\documentclass[12pt]{article}
\usepackage{amsmath}
\usepackage{epsfig,graphics}

\newcommand{\be}{\begin{equation}}
\newcommand{\ee}{\end{equation}}

\newlength{\figsize}
\begin{document}

\begin{titlepage}

\vspace*{0.8in}
 
\begin{center}
{\large\bf Confinement and the effective string theory
\\ in SU($N\to\infty$) : a lattice study.\\ }
\vspace*{0.75in}
{Harvey Meyer$^{a}$ and Michael Teper$^{b}$\\
\vspace*{.45in}
$^{a}$DESY Zeuthen, Platanenallee 6, 15738 Zeuthen, Germany\\
\vspace*{.05in}
$^{b}$ Rudolf Peierls Centre for Theoretical Physics, University of Oxford,\\
1 Keble Road, Oxford OX1 3NP, U.K.
}
\end{center}

\vspace*{0.75in}

\begin{center}
{\bf Abstract}
\end{center}

We calculate in the SU(6) gauge theory the mass of the 
lightest flux loop that winds around a spatial torus, as
a function of the torus size, taking care to achieve control 
of the main systematic errors. For comparison we perform a 
similar calculation in SU(4). We demonstrate 
approximate linear confinement and show that the leading
correction is consistent with what one expects if the
flux tube behaves like a simple bosonic string at long
distances. We obtain similar but less accurate results for
stable ($k$-)strings in higher representations. We find
some evidence that for $k>1$ the length scale at 
which the bosonic string correction becomes dominant 
increases as $N$ increases. We perform all these 
calculations not just for long strings, up to about 2.5`fm'
in length, but also for shorter strings, down to the minimum 
length, $l_c = 1/T_c$, where $T_c$ is the 
deconfining temperature. We find that the mass of the 
ground-state string, at all length scales, is not very far from 
the simple Nambu-Goto string theory prediction,
and that the fit improves as $N$ increases from $N=4$ to $N=6$. 
We estimate the mass of the first excited string and find that 
it also follows the Nambu-Goto prediction, albeit more
qualitatively. We comment upon the significance of these results 
for the string description of SU($N$) gauge theories in the 
limit $N=\infty$.

\end{titlepage}

\setcounter{page}{1}
\newpage
\pagestyle{plain}

\section{Introduction}
\label{section_intro}

There is considerable numerical evidence for linear confinement
in SU(2) and SU(3) gauge theories and significant
evidence that the effective string theory 
governing the long-distance dynamics of confining flux tubes 
is in the same universality class as the simplest Nambu-Goto 
bosonic string. (See
\cite{blmt-string01,luscher-string02,other-string}
for some recent calculations.) For SU($N>3$) gauge theories,
while there is some evidence for linear confinement
\cite{blmt-string01,blmt-string00},
these questions have not been addressed with much precision.
In particular, this is so for the stable $k$-strings in higher
representations which have been the object of intensive recent
study 
\cite{blmt-string01,blmt-string00,pisa-kstring,blmtuw-glue04}.
One usually assumes a simple bosonic string 
correction for these $k$-strings, and one also assumes that 
this leading correction dominates on the same length
scales as it does for strings in the fundamental representation 
in SU(2) and SU(3). 
Since the comparison with theoretical predictions for
$k$-string tensions is sensitive to this assumption, it is
important to check its validity. In addition not much
is known about the effective string theory at large $N$.
Since it is in the $N=\infty$ limit that the gauge theory
might be equivalent to a string theory 
\cite{largeN-string}
and since it
is in that limit that one can currently make quantitative
contact between field theories and their string theory duals
\cite{largeN-ads},
it is important to learn what are the properties of the
effective confining string theory in that limit.

In this paper we address these questions with a lattice calculation
in the SU(6) gauge theory. From other work on the mass spectrum
and the deconfining temperature
\cite{blmt-glue01,blmtuw-glue04,hmmt-regge04}
we know that $N=6$ is very close to $N=\infty$ for many
quantities. This is no surprise since the leading large-$N$ 
correction is expected to be $O(1/N^2)$. Our calculations
are at a fixed value of the lattice spacing $a$. In units
of the fundamental string tension $\sigma$, the value of $a$ 
is $a \simeq 1/4\surd\sigma$ which is well into the weak
coupling region where any corrections to continuum physics
should be small. For comparison we also perform a calculation
in the SU(4) gauge theory at a similar value of $a$.

The main questions we address are as follows. \\
$\bullet$ Do we have linear confinement at large $N$?
To address this we calculate the energy of a
`flux tube' that winds around a spatial torus. We  
then vary the length of that torus and see if the energy grows 
(approximately) linearly with the length, once the
length is large.  \\
$\bullet$ What is the leading correction to this linear
behaviour at large $l$? Equivalently: what is the central
charge of the effective string theory that describes
the long distance confining physics?  \\
$\bullet$ If we reduce the length of the flux tube 
-- indeed, to its minimum value -- 
how does its energy compare to the behaviour predicted
by the simple Nambu-Goto action? This addresses the
question of what is the effective string theory on all
scales. In particular, do things become
simpler as $N\to\infty$?  \\
$\bullet$ How does the energy of the first excitation
of the string compare to the Nambu-Goto prediction?
This probes the content of the effective string theory
in a different way.  \\
$\bullet$ How does all this affect the reliability of
our calculations of $k$-string tensions? And indeed,
what is our best estimate of these tensions?

In the next Section we shall describe in greater detail
the background to these questions. We then summarise
the technical details of the lattice calculation.
We then move on to the calculation itself. We begin
by quantifying what are potentially the most serious
systematic errors in such calculations. We then 
interpret our results. We finish with a summary of what 
we have learned and then describe the next step in improving
upon these calculations.

\section{Preliminaries}
\label{section_preliminaries}

We shall begin with a discussion of how our calculations
of the string energy as a function of length can
probe the nature of the effective string theory.
We then briefly outline the lattice setup of the calculations.

\subsection{Strings}
\label{subsection_strings}

If we have linear confinement, we expect that the mass
of a flux loop that winds around a spatial torus of length $l$ 
will be given by
\cite{gsmt-string84,luscher-string80}
\begin{equation}
m(l) 
\stackrel{l\to\infty}{=}
\sigma l - \frac{\pi}{3} \frac{c}{l}
\label{eqn_corr} 
\end{equation}
where $c$ is the central charge of the effective string 
theory describing the long-distance properties of the
confining flux tube. (The dimensional factor, $d-2$,
has been included in the coefficient.) 
There is significant numerical evidence that $c=1$
for SU(2) and SU(3) gauge theories
\cite{blmt-string01,luscher-string02,other-string},
corresponding to the universality class of the
simplest bosonic string. In this paper we shall
attempt to determine the value of $c$ for SU(6),
in the expectation that this will also be its value at
$N=\infty$.

There are old ideas that in its confining phase the 
SU($N=\infty$) gauge theory might be a string theory.
It is therefore interesting to calculate $m(l)$ not 
just for large $l$ but for all $l$, and indeed for excited 
strings as well. This is also interesting in the 
context of the more recent dual string approaches:
while $c$ in eqn(\ref{eqn_corr}) determines the number
of exactly massless modes in the string theory, there
will be massive modes generated 
in the descent to a boundary SU($\infty$) gauge
theory, which may be encoded, for example, in the excited 
state spectrum of the string.

The very simplest bosonic string theory in $d$ space-time
dimensions, based on the Nambu-Goto action, has an energy spectrum
\cite{arvis-string83,luscher-string04}
\begin{equation}
E_n(l) 
 = 
\sigma l 
\Bigl\{ 
1 + \frac{8\pi}{\sigma l^2}\bigl( n -  \frac{d-2}{24}\bigr)
\Bigr\}^\frac{1}{2}
\label{eqn_NGE} 
\end{equation}
and a ground state energy, in d=4, 
\begin{equation}
E_0(l) 
\stackrel{l\to\infty}{\longrightarrow} 
\sigma l - \frac{\pi}{3} \frac{1}{l}
- \frac{\pi^2}{18\sigma} \frac{1}{l^3} + \ldots
\label{eqn_NG} 
\end{equation}
The  correction to the universal $O(1/l)$ term in 
$E_0(l)$ is $O(1/l^3)$ and, as one would expect,
the expansion is in inverse powers of $l^2\sigma$.
This is consistent with the observation in
\cite{blmt-string01}
that the universal $O(1/l)$ term is the dominant correction
for $l\surd\sigma \geq 3$. If one adds the most general 
higher-dimensional operators to the free term in the action, 
while remaining within an effective string theory with the 
same degrees of freedom, then one can show
\cite{luscher-string04}
that the expansion in eqn(\ref{eqn_NG}) retains its form except
that the coefficient of the $O(1/l^3)$ term is now undetermined.
If one includes in addition to interactions amongst the 
massless transverse modes, massive degrees of freedom 
which naturally arise in dual string constructions,
then these would typically be encoded in excited states
whose energies do not approach $E_0(l)$ as $l\to\infty$
\cite{ofer}.
Thus we can expect the string spectrum to contain interesting and
useful information about the details of the equivalent string theory.
In this spirit we will calculate the energy of the lightest and first 
excited strings as a function of $l$.

Of course, as one decreases $l$ one encounters a minimum length for 
which a winding flux loop exists. This is easy to see. Suppose we
choose the flux loop to wind around the $x$-torus. Now imagine
relabelling co-ordinates $x,y,z,t \to t,x,y,z$. In this case
the length $l$ equals the inverse temperature and 
the mass of the winding string gives the effective string
tension (times $l$)
\cite{blmtuw-T03}.
If $T_c$ is the deconfining temperature then for $T>T_c$
there is no confinement and no winding string. Thus
the minimum value of $l$ is $l_c =1/T_c$. 

A linearly confining theory naturally embodies deconfinement 
because the number of strings of length $l$ grows as 
$\propto \exp\{+\gamma l\}$ with $\gamma >0$ and this will clearly 
overwhelm the Boltzman suppression $ \propto \exp\{-\sigma l/T\}$
once $T$ is large enough
\cite{polyakov}.
This argument naturally suggests a vanishing effective string
tension at $T=T^s_c$ i.e. a second order transition. As we see
from eqn(\ref{eqn_NGE}), this occurs in the Nambu-Goto 
action at the length $l=1/T^s_c$ where $E_0(l)$ vanishes, i.e. at
\begin{equation}
\frac{T^s_c}{\surd\sigma} 
=
\sqrt{\frac{3}{(d-2)\pi}}
\simeq \begin{cases} 0.691 &:\quad d=4\\
0.977  &:\quad d=3
\end{cases}
\label{eqn_NGTc} 
\end{equation}
Of course this string deconfining temperature might be
trumped by a transition to the gluon plasma at a lower
value of $T=T_c < T^s_c$, driven by the large entropy
of that plasma. In this case it is natural 
for the transition to be first order. It is therefore
interesting to compare the predicted values in
eqn(\ref{eqn_NGTc}) to the calculated values 
of $T_c$ in those cases where the transition is in fact
second order. One finds
\cite{blmtuw-T03,mt-d3T}
\begin{equation}
\frac{T_c}{\surd\sigma} 
=
\begin{cases} 0.709(4) &:\quad SU(2), d=4\\
1.12(1)  &:\quad SU(2), d=3\\
0.98(2)  &:\quad SU(3), d=3
\end{cases}
\label{eqn_LatTc} 
\end{equation}
We note a quite remarkable co-incidence with the Nambu-Goto 
values. (This remark is not new: see e.g.
\cite{oldTc-NG}.)
Of course what we are really interested in is the 
$N\to\infty$ limit, where there are theoretical arguments that 
the gauge theory might be equivalent to a string theory,
and here deconfinement is known to be first order
\cite{blmtuw-T03}.
Nonetheless this provides us with a bound
\begin{equation}
\lim_{N\to\infty} \frac{T^s_c}{\surd\sigma} 
>
\lim_{N\to\infty} \frac{T_c}{\surd\sigma} 
=
0.596(4)
\ \ \ \; \ \ \ d=4.
\label{eqn_NooTc} 
\end{equation}

The Nambu-Goto string is of course extremely simple -- no
doubt much too simple. From string and dual string approaches
\cite{ofer}
one expects 
\begin{equation}
\frac{T^s_c}{\surd\sigma} 
\propto
\frac{1}{\surd c}
\label{eqn_NooTc2} 
\end{equation}
where $c$ counts the number of effective degrees of freedom
in the string theory, on the scale of $T_c$. Thus one
typically would expect a value of $T^s_c/\surd\sigma$ that
is lower than the Nambu-Goto value. And if one could
obtain $\lim_{N\to\infty} T^s_c/\surd\sigma$ one might learn
something interesting about the dual string theory.
It might seem that this limit is inaccessible given that
the transition is first order for all $N\geq 3$. However this 
is not necessarily so
\cite{mt-unpub}. 
If the tension of the interface between confined and 
deconfined phases is $\propto N^2$ (its natural dependence)
then the hysteresis around $T=T_c$ can become large
\cite{blmtuw-T03,blmtuw-T04}
on even moderately large spatial volumes.
That is to say, if we are in the confined phase we will
stay in that phase as we increase $T$ above $T_c$ and  even 
as $T\to T^s_c$, if the hysteresis is extended enough.
This is a realistic possibility. Preliminary calculations
\cite{mt-unpub}
in SU(8) indicate that for $N=8$ the hysteresis is not (yet?) 
wide enough. Nonetheless the calculation provides an improved 
bound of $T^s_c/\surd\sigma > 0.63$. 

One can extend the above discussion from strings in the
fundamental representation to strings in higher representations
\cite{blmt-string01}.
It is useful to classify strings by how their sources transform 
under a gauge transformation belonging to the centre, $z\in Z_N$ 
for SU($N$), since gluon screening does not transform sources
between these classes. Let the transformation be $z^k$. Then
the lightest string in this class is called the $k$-string.
(The usual string is thus a 1-string.) 
One can think of it as a collection of $k$ fundamental strings 
and an interesting question is whether they form a bound string 
state whose tension, $\sigma_k$, is less than the $k$ separate 
strings, i.e. $\sigma_k < k\sigma$. It is also 
easy to see that $k$ is less than the integer part of $N/2$
so that we have to go to at least SU(4) to have the possibility
of this new type of string. In fact this possibility is realised
\cite{blmt-string01}:
the $k$-strings are strongly bound and the spectrum has been
calculated quite accurately for $k\leq 4; N\leq 8$
\cite{blmtuw-glue04}. 
Since the Nambu-Goto string action has no interactions, it cannot
give such a binding and so cannot be the whole story at finite $N$.
However as $N\to\infty$ the binding disappears,  
$\sigma_k \to k\sigma$, and the Nambu-Goto
action re-emerges as a possibility. One might also entertain the
possibility of the deconfining transition occurring in discrete
steps at different temperatures, with the $k$-string deconfining
at $T=T^k_c$, but it turns out that in practice this does not
occur -- there is a single first order transition for $N\geq 4$
\cite{blmtuw-T04}. 
And one can also argue that the same is the case for the would-be
string deconfining temperatures $T^{s,k}_c$
\cite{blmtuw-T04}.

The calculation of $\sigma_k$ requires calculating the mass
of a $k$-string of length $l$ and then applying a correction 
as in eqn(\ref{eqn_corr}). Typically one assumes that this
correction is the same as for the fundamental string, $c=1$,
and that it is the only important correction for 
$l\surd\sigma \geq 3$ because that is what one finds in SU(2) 
and SU(3) for the fundamental string. This assumption is 
important at the level of accuracy required to distinguish 
between competing theoretical ideas about the $k$-strings.
Thus one of the things we will do is to try to obtain
some direct numerical evidence with accurate calculations
in SU(6) for the $l$-dependence of the masses of the
lightest $k=1,2,3$ strings.

\subsection{Lattice}
\label{subsection_lattice}

We work on periodic hypercubic $L_x\times L_y\times L_z\times L_t$ 
lattices with lattice spacing $a$. 
The degrees of freedom are SU($N$) matrices, 
$U_l$, defined on the links $l$ of the lattice. The
partition function is 
\begin{equation}
{\cal Z}(\beta)
=
\int \prod_l dU_l 
e^{-\beta \sum_p\{ 1 - \frac{1}{N}{\mathrm{ReTr}}u_p\}}
\ \ \ \ ; \ \ \ \
\beta=\frac{2N}{g^2}
\label{eqn_lattice} 
\end{equation} 
where $u_p$ is the ordered product of matrices around the
boundary of the elementary square (plaquette) labelled by $p$
and $g^2$ is the bare coupling. This is the standard
Wilson plaquette action and since the theory is asymptotically
free and since the bare coupling is a
running coupling on length scale $a$, the continuum limit is 
approached by tuning $\beta = 2N/g^2(a) \to \infty$. 
One expects that for large $N$ the value of $a$ is fixed  
in physical units (e.g. in units of the mass gap) if one
keeps the 't Hooft coupling $\lambda(a) \equiv g^2(a)N$
fixed i.e. $\beta \propto N^2$. This has been confirmed
\cite{blmt-glue01}
in non-perturbative lattice calculations. 

We will consider a loop of flux that winds around the 
$x$-torus, so that it is of length $l=aL_x$. 
(Our notation is that $l$ denotes a length in either physical 
or lattice units, depending on the context, while 
the capitalized form is always in lattice units.)
A generic operator
$\phi_l$ that couples to such a periodic flux loop is an 
ordered product of link matrices along a space-like curve 
that winds once around the $x$-torus. 
Correlations are taken in the $t$ direction so that
the energy of the loop is an eigenstate of the Hamiltonian
(transfer matrix) defined on the $xyz$ space. 
Such a correlator may be expanded 
\begin{equation}
C(t=an_t) 
\equiv \langle \phi^\dagger_l(t) \phi(0) \rangle
=
\sum_{n=0}  |\langle n |\phi(0) | vac \rangle|^2 e^{-aE_n n_t}
\ \ \ \ ; \ \ \ \
E_i \leq E_{i+1}
\label{eqn_corrln} 
\end{equation} 
where, if we have confinement, $E_0$ is the energy of the 
lightest flux loop. Since the fluctuations that determine the 
error in the Monte Carlo calculation of $C(t)$ may be expressed
as a correlation function with a disconnected piece, the
error is approximately independent of $t$ while $C(t)$ itself
decreases exponentially with $t$. Thus one needs operators
$\phi_l$ that have very large projections onto the
desired state(s) so that this state dominates $C(t)$ at small $t$. 
This can be achieved using now standard techniques (see e.g.
\cite{blmtuw-glue04}).
If we have linear confinement $E_0$ becomes large for long loops,
and especially so for higher representations, and $C(t)$ becomes
very small. It can then be difficult, even at small $t$, to make 
the error on $C(t)$ much smaller than its value, if one
uses a standard Monte Carlo.
In some of these cases we therefore use a recently developed
error reduction technique
\cite{hm03}.

We choose $L_y=L_z$ for convenience. We need to make 
sure that $L_y$ is large 
enough for there to be no significant finite volume effects.
We will also need to choose $L_t$ large enough so that the only
state with a significant probability to propagate around the
$t$-torus is the vacuum state. In practice it turns out that
for large $L_x$ one can use  $L_y=L_z=L_t=L_x$, but for
small  $L_x$ we shall need to use  $L_y, L_t\gg L_x$.
Controlling these different finite volume effects is one
way in which the present calculation improves upon previous
calculations. 

Our calculations will be performed for a single lattice 
spacing. In the case of SU(6) we choose $\beta = 25.05$.
This corresponds to 
\begin{equation}
a(\beta=25.05) 
= 
\frac{0.252(1)}{\surd\sigma}
=
\frac{0.1509(5)}{T_c}
\ \ \ \ : \ \ \ \
SU(6)
\label{eqn_scaleN6} 
\end{equation} 
in units of the fundamental string tension $\sigma$ (see
below) or the deconfining temperature $T_c$
\cite{blmtuw-T03}. 
In SU(4) we choose $\beta=10.90$, which corresponds to 
\begin{equation}
a(\beta=10.9) 
= 
\frac{0.241(2)}{\surd\sigma}
=
\frac{0.1502(6)}{T_c}
\ \ \ \ : \ \ \ \
SU(4).
\label{eqn_scaleN4} 
\end{equation} 
The value of the lattice spacing has been chosen to be
(almost) the same for SU(4) and SU(6) so that we can separate
the $N$-dependence from the $a$-dependence (to leading order).
This value of $a$ is small enough that we can expect
lattice corrections to be negligible for our purposes.

\section{Results}
\label{section_results}

We calculate the string mass as a function of 
its length on a variety of lattice volumes. In Table~\ref{table_su6}
we list the masses obtained in SU(6) at $\beta=25.05$
for the lightest $k=1,2,3$ strings and, on the largest
lattices, for the first excited $k=1$ string.
In Table~\ref{table_su4} we list a less extensive set of results 
obtained in SU(4) at $\beta=10.9$. 

We will begin by describing the important sources of 
systematic error in such calculations and we show how the 
results listed in Table~\ref{table_su6} enable us to control 
these errors. Having obtained the string energies $E^k_0(l)$ for
$l$ ranging from close to its minimum possible value upwards,
we address the questions posed in the Introduction.

\subsection{Systematic errors}
\label{subsection_errors}

There are a large number of potential systematic errors.
Here we focus on three that we believe are potentially
the most important ones in this kind of calculation.

\subsubsection{excited state contamination}
\label{subsection_contamination}

Our variational calculation produces operators $\Phi^k_l(t)$ that
have very good but not perfect overlaps onto the ground string
states. We therefore calculate the correlator 
$C(t) = \langle \Phi^{k\dagger}_l(t)\Phi^k_l(0)\rangle$ 
and extract the energy from the larger values of $t=an_t$ where 
the excited state contamination has died out. The signal
that we have reached such $t$ is that the correlator is given by 
a single exponential. A problem arises if  $E^k_0(l)$ is large --
either because $l$ is large or because $k$ is large.
In this case the correlator will be very small beyond the 
smallest values of $t$, so that it becomes dominated by the 
statistical error and one cannot obtain
any significant evidence that the higher excited states 
have in fact died away at the $t$ at which one extracts the energy.
Clearly the bias will be to extract energies at lower $t$
as the energy increases and so to risk a larger admixture of 
even heavier excited states, and so to overestimate the energy
as $k$ or $l$ increases. This risk is enhanced for larger $k$
because the best overlap decreases roughly as the $k$'th power
of the $k=1$ overlap. (Which provides a motivation for our
description of the $k$-string as a loosely bound state of
$k$ fundamental strings.) It is also enhanced for large $l$
by two facts. First the gap between ground and excited
states decreases as $\propto 1/l$ and second the overlap
onto the lightest string appears to decrease exponentially
with its length $l$. The second effect is not unexpected
and is in practice not severe: the overlap reduces from
$\sim 0.985$ for $l=8$ to $\sim 0.960$ for $l=16$ using 
a very similar basis of operators. 

All this is not an issue for the smallest values of $l$
where we have accurate calculations of $C(t)$ over a large
range of $t$. To obtain some control of this error for 
larger $l$, in particular for $l=16$ and $l=20$, we have 
performed not only a normal Monte Carlo calculation of
the correlator, but also a separate calculation using a 
novel and powerful error reduction technique
\cite{hm03,luscher01}.  
The latter calculations are starred in Table~\ref{table_su6}.
The error reduction technique works best for a particular mass
range for which it needs to be tuned in advance. Here we have
tuned the calculation to work best for masses close to those
of the lightest $k=2$ string.
The unstarred calculations use a standard Monte Carlo algorithm 
and, although the masses are extracted 
from an exponential fit to $C(t)$ between $t=a$ and $t=4a$,
in the case of $l=16$ and $l=20$
the signal-to-error ratio grows so rapidly with increasing $t$
that, effectively, the masses are obtained from $t=a$ to $t=2a$.
The starred calculations involve extracting the masses from
$t=2a$ to $t=3a$ where we can be quite confident that any
contamination from heavier excited string states will be
negligible.  

Comparing the results from the starred and unstarred calculations
in  Table~\ref{table_su6} we find very good agreement within
the pairs of calculations for all the states. This provides good
direct evidence that any contamination from excited states is
negligible in the present calculation.

\subsubsection{transverse size corrections}
\label{subsection_transvers}

Calculations of winding flux loops have usually been performed 
on $L^3L_t$ lattices. Is a transverse size of $L$ large enough
for a flux tube of length $L$ or does this create significant
finite volume corrections? To the extent that $L$ is large
compared to the width of the flux tube, and the oscillations
are simple single-valued harmonic modes, 
one would expect the corrections to be very small. So unlike 
the systematic error discussed above, this is expected
to affect short rather than long strings. And to the extent
that a $k$-string is like $k$ weakly bound $k=1$
strings, we might expect any error to be more severe
for larger $k$.  

To address this question we start by comparing the three
calculations in
Table~\ref{table_su6} that have $L_x=8$ and $L_t=30$.
(The corrections induced by too small a value $L_t$
will be discussed below.) These three calculations have 
transverse sizes  $L_y=L_x=8$, $L_y=10$ and  $L_y=12$.
We observe that there is no discrepancy in the masses obtained
with the last two, implying that for $L_x=8$ a transverse size
of $L_y=10$ is in fact large enough. The $L_y=L_x=8$
calculation, on the other hand, produces masses that are very 
different: while for $k=1$ the shift is a modest $\simeq 10\%$,
for $k=2,3$ the shifts are $\sim 24\%$ and $\sim 35\%$
respectively. Our expectation that the corrections should 
be larger for larger $k$ thus appears to be confirmed.

Moving to $L=10$ we have $L_y=10$ and $L_y=12$ to compare. For 
$k=1$ there appears to be no difference, while there is evidence
for a modest discrepancy $\sim 5\%$ for the $k=2$ string.
That this is not merely a statistical fluctuation is confirmed
if we compare the $k=2$ effective masses obtained from the 
correlator between $t=0$ and $t=a$, where the relative statistical
error is much smaller. The conclusion is that for $L=10$ there
is a shift in using $L_y=10$ but that it is very small. Moving
on to $L=12$ we compare $L_y=12$ and $L_y=14$. The lightest and
first excited $k=1$ strings show no difference and the possible
difference for the $k=2,3$ strings is shown to be no more than
a statistical fluctuation when we look at the statistically much 
more accurate $t=0,a$ effective masses.

We therefore conclude that, for this value of $a$, once $L\geq 12$ 
it is safe to use an $L^3L_t$ lattice for such string calculations.
Using $a\surd\sigma \simeq 0.25$ this translates into the 
statement that if the length $l=aL_x$ satisfies 
$l\surd\sigma \geq 3$ then an equal transverse size is adequate.
It so happens that, for quite different reasons, lattices used
for calculating $k$-string tensions have satisfied precisely
this criterion.

\subsubsection{temporal size corrections}
\label{subsection_temporal}

If all the spatial dimensions, $L_i$, are large then all 
the string energies will be large, and $\exp\{-EL_t\}$ 
will be negligibly small for all states other 
than the vacuum if we choose, for example, $L_t=L_i$.
(Recall that all glueball masses will in practice be large
compared to $1/L_t$.) So for larger $L_i$ the partition function
is indeed dominated by the vacuum and there is no problem.
For small $L$, however, we can only be sure of excluding 
the spatially periodic strings from propagating right around 
the temporal torus if we choose $L_t$ so that
$\exp\{-E_0(L)L_t\}\ll 1$. This means that we must increase
$L_t$ as we decrease $L$. If we do not do so, then the
partition function, which is a normalising factor
within all our calculated expectation values, will no longer 
dominated by the vacuum contibution and we run the risk that 
our various calculated energies will 
no longer be with respect to the correct vacuum energy.

In practice one finds that for the $k=1$ string such an
effect is quite weak -- as we see by comparing the
$8^3 14$ and $8^3 30$ lattices in Table~\ref{table_su6}
or the $10^4$ with the $10^3 16$. This is as expected:
e.g. for the $8^3 14$ lattice one sees that
$\exp\{-E_0(L)L_t\} \sim 0.015$ and there is a factor of
3 for the strings  in the different spatial directions.
So the expected shift in the effective vacuum energy is only
$\Delta E_{vac} \sim 0.003$. In fact it is well known
that even with smaller $L_t$ the corrections are smaller
than expected, and this can be ascribed to a cancellation 
between these extra modes in the path integral numerator
and in normalising denominator (partition function) 
in the calculation of an expectation value. One would expect
such a cancellation if the interaction between 
this extra mode and the string propagating between the
operators of the correlator was weak. 

However, as we see in Table~\ref{table_su6}, there is a quite 
dramatic mass shift for the $k=2$ and $k=3$ masses when 
we compare, for example, the $8^3 14$ and $8^3 30$ lattices.
We can interpret this, in part, as follows. Consider
our $k=2$ calculation. We
have two $k=2$ operators separated by $t$ and we assume
that for large enough $t$ the correlator is dominated by 
the lightest $k=2$ loop propagating between these operators.
However an alternative contribution arises from two
$k=1$ loops propagating between the operators in
opposite directions around the temporal torus. Such 
a contribution is $t$-independent and acts, in the
calculation, like an unexpected  non-zero
vacuum expectation value for the $k=2$ loop operator.
This will lower the calculated effective mass at larger $t$.
It is therefore clear from this that we require $L_t$ to
be large enough that 
$\exp\{-E^{k=1}_0(L)L_t\} \ll  \exp\{-E^{k}_0(L)t\}$
for any value of $t$ at which we might evaluate the
mass of the $k$-string.

\subsection{Interpretation}
\label{subsection_interpret}

Our above discussion of systematic errors suggests that
the following choice of lattices from  Table~\ref{table_su6}
will provide reliable string masses. For $l = 7,8,10$ we use
the $7\times 16^2 \times 40$, $8\times 12^2 \times 30$ and
, $10 \times 12^2 \times 16$ lattices respectively.
For $l\geq 12$ we average over the pairs of calculations
at each $l$.

\begin	{figure}[p]
\begin	{center}
\leavevmode
\input	{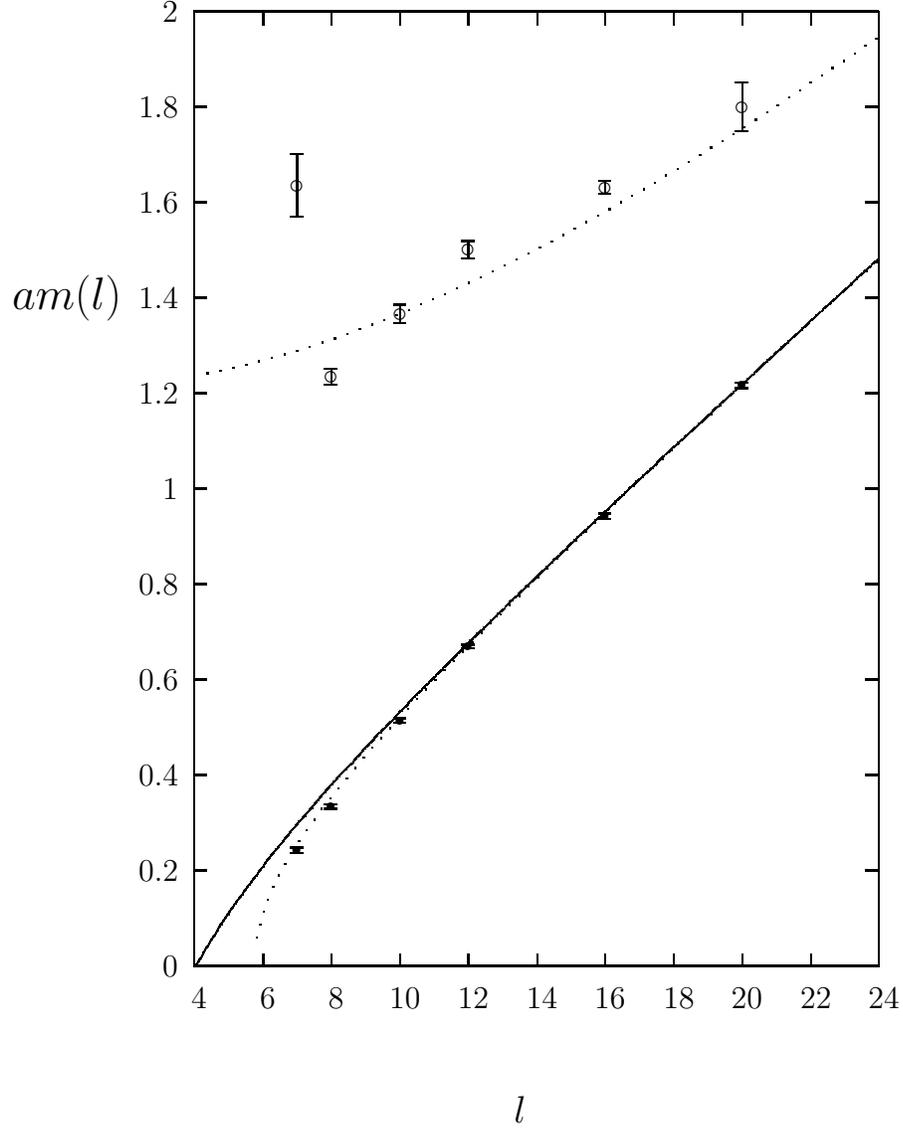}
\end	{center}
\vskip 0.15in
\caption{The masses of the lightest, $\bullet$, and first
excited, $\circ$, $k=1$ flux loops that wind
around a spatial torus of length $l$ in the SU(6) 
calculation at $\beta=25.05$. The dotted
lines are the predictions of the Nambu-Goto string action,
as in eqn(\ref{eqn_NGE}). The dynamical lower bound on the 
string length is $l_{min} = 1/aT_c \simeq 6.63$.}
\label{fig_K1}
\end 	{figure}

In Fig.\ref{fig_K1} we plot the lightest and first excited
string masses versus their length $l$. The first thing we 
observe is that the lightest loop mass, $E_0(l)$, grows 
more-or-less linearly with its length, $l$. But before 
claiming that this provides  direct evidence for linear 
confinement we need to establish that this is not some
sub-asymptotic behaviour at short distances. Now,
if we extract a string tension (see below) we obtain 
$a\surd\sigma \simeq 0.25$. Thus our largest flux loop length 
is $20a \simeq 5/\surd\sigma \simeq 2.5$`fm'.
(The `fermi' units are obtained using the QCD value
of the string tension and are only introduced to provide
an intuitively familiar scale of length.) This is long
enough to make it plausible that we are indeed seeing the asymptotic
linear rise of a linearly confining SU(6) gauge theory. 
Since we expect corrections to the $SU(\infty$) theory
to be $O(1/N^2)$ and since the extracted value of the
string tension is very similar to that in, for example,
SU(3) when calculated in units of, for example, the mass gap
\cite{blmt-glue01,blmtuw-glue04},
this provides very plausible evidence for linear confinement in
the $N=\infty$ gauge theory.

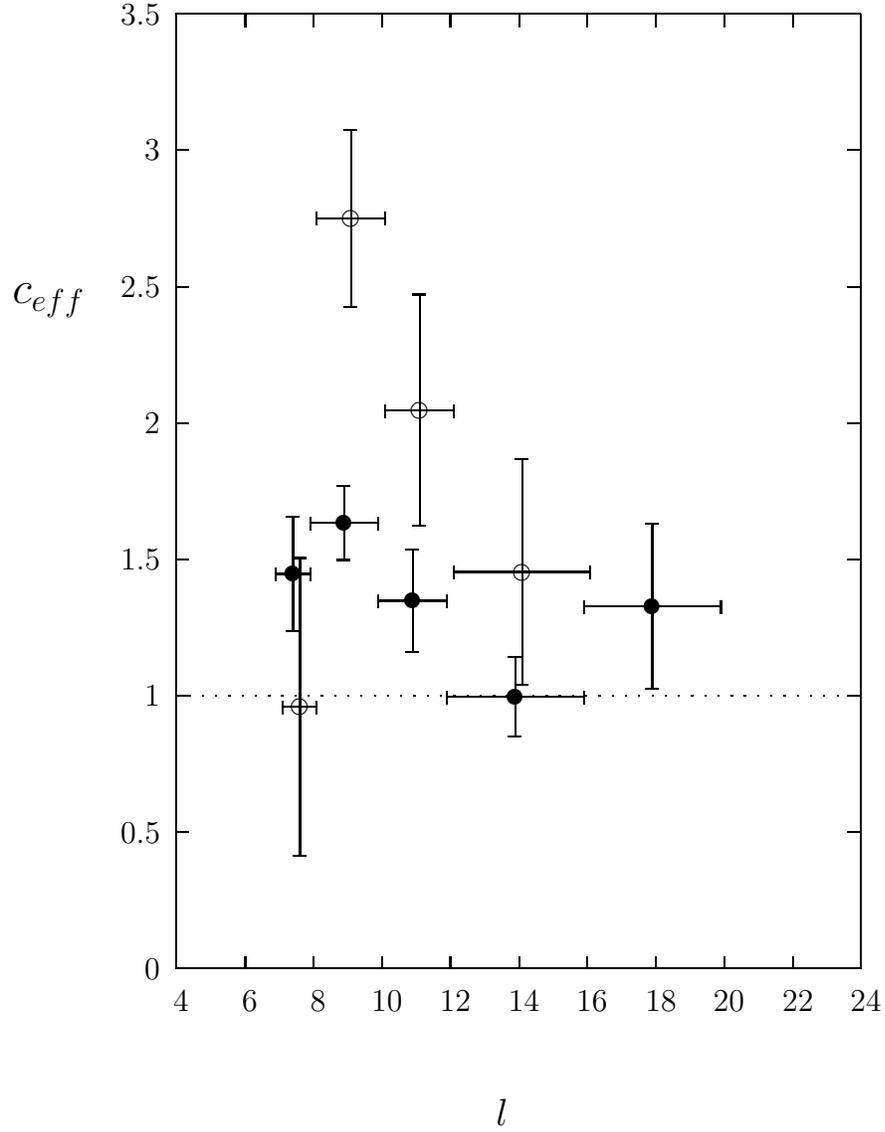
\begin	{figure}[p]
\begin	{center}
\leavevmode
\setlength{\unitlength}{0.240900pt}
\ifx\plotpoint\undefined\newsavebox{\plotpoint}\fi
\sbox{\plotpoint}{\rule[-0.200pt]{0.400pt}{0.400pt}}%
\begin{picture}(1500,1800)(0,0)
\font\gnuplot=cmr10 at 12pt
\gnuplot
\sbox{\plotpoint}{\rule[-0.200pt]{0.400pt}{0.400pt}}%
\put(350.0,250.0){\rule[-0.200pt]{4.818pt}{0.400pt}}
\put(325,250){\makebox(0,0)[r]{\ \ {$0$}}}
\put(1405.0,250.0){\rule[-0.200pt]{4.818pt}{0.400pt}}
\put(350.0,464.0){\rule[-0.200pt]{4.818pt}{0.400pt}}
\put(325,464){\makebox(0,0)[r]{\ \ {$0.5$}}}
\put(1405.0,464.0){\rule[-0.200pt]{4.818pt}{0.400pt}}
\put(350.0,679.0){\rule[-0.200pt]{4.818pt}{0.400pt}}
\put(325,679){\makebox(0,0)[r]{\ \ {$1$}}}
\put(1405.0,679.0){\rule[-0.200pt]{4.818pt}{0.400pt}}
\put(350.0,893.0){\rule[-0.200pt]{4.818pt}{0.400pt}}
\put(325,893){\makebox(0,0)[r]{\ \ {$1.5$}}}
\put(1405.0,893.0){\rule[-0.200pt]{4.818pt}{0.400pt}}
\put(350.0,1107.0){\rule[-0.200pt]{4.818pt}{0.400pt}}
\put(325,1107){\makebox(0,0)[r]{\ \ {$2$}}}
\put(1405.0,1107.0){\rule[-0.200pt]{4.818pt}{0.400pt}}
\put(350.0,1321.0){\rule[-0.200pt]{4.818pt}{0.400pt}}
\put(325,1321){\makebox(0,0)[r]{\ \ {$2.5$}}}
\put(1405.0,1321.0){\rule[-0.200pt]{4.818pt}{0.400pt}}
\put(350.0,1536.0){\rule[-0.200pt]{4.818pt}{0.400pt}}
\put(325,1536){\makebox(0,0)[r]{\ \ {$3$}}}
\put(1405.0,1536.0){\rule[-0.200pt]{4.818pt}{0.400pt}}
\put(350.0,1750.0){\rule[-0.200pt]{4.818pt}{0.400pt}}
\put(325,1750){\makebox(0,0)[r]{\ \ {$3.5$}}}
\put(1405.0,1750.0){\rule[-0.200pt]{4.818pt}{0.400pt}}
\put(350.0,250.0){\rule[-0.200pt]{0.400pt}{4.818pt}}
\put(350,200){\makebox(0,0){\ {$4$}}}
\put(350.0,1730.0){\rule[-0.200pt]{0.400pt}{4.818pt}}
\put(458.0,250.0){\rule[-0.200pt]{0.400pt}{4.818pt}}
\put(458,200){\makebox(0,0){\ {$6$}}}
\put(458.0,1730.0){\rule[-0.200pt]{0.400pt}{4.818pt}}
\put(565.0,250.0){\rule[-0.200pt]{0.400pt}{4.818pt}}
\put(565,200){\makebox(0,0){\ {$8$}}}
\put(565.0,1730.0){\rule[-0.200pt]{0.400pt}{4.818pt}}
\put(673.0,250.0){\rule[-0.200pt]{0.400pt}{4.818pt}}
\put(673,200){\makebox(0,0){\ {$10$}}}
\put(673.0,1730.0){\rule[-0.200pt]{0.400pt}{4.818pt}}
\put(780.0,250.0){\rule[-0.200pt]{0.400pt}{4.818pt}}
\put(780,200){\makebox(0,0){\ {$12$}}}
\put(780.0,1730.0){\rule[-0.200pt]{0.400pt}{4.818pt}}
\put(888.0,250.0){\rule[-0.200pt]{0.400pt}{4.818pt}}
\put(888,200){\makebox(0,0){\ {$14$}}}
\put(888.0,1730.0){\rule[-0.200pt]{0.400pt}{4.818pt}}
\put(995.0,250.0){\rule[-0.200pt]{0.400pt}{4.818pt}}
\put(995,200){\makebox(0,0){\ {$16$}}}
\put(995.0,1730.0){\rule[-0.200pt]{0.400pt}{4.818pt}}
\put(1103.0,250.0){\rule[-0.200pt]{0.400pt}{4.818pt}}
\put(1103,200){\makebox(0,0){\ {$18$}}}
\put(1103.0,1730.0){\rule[-0.200pt]{0.400pt}{4.818pt}}
\put(1210.0,250.0){\rule[-0.200pt]{0.400pt}{4.818pt}}
\put(1210,200){\makebox(0,0){\ {$20$}}}
\put(1210.0,1730.0){\rule[-0.200pt]{0.400pt}{4.818pt}}
\put(1318.0,250.0){\rule[-0.200pt]{0.400pt}{4.818pt}}
\put(1318,200){\makebox(0,0){\ {$22$}}}
\put(1318.0,1730.0){\rule[-0.200pt]{0.400pt}{4.818pt}}
\put(1425.0,250.0){\rule[-0.200pt]{0.400pt}{4.818pt}}
\put(1425,200){\makebox(0,0){\ {$24$}}}
\put(1425.0,1730.0){\rule[-0.200pt]{0.400pt}{4.818pt}}
\put(350.0,250.0){\rule[-0.200pt]{258.967pt}{0.400pt}}
\put(1425.0,250.0){\rule[-0.200pt]{0.400pt}{361.350pt}}
\put(350.0,1750.0){\rule[-0.200pt]{258.967pt}{0.400pt}}
\put(150,1300){\makebox(0,0){\Large{$c_{eff}$}}}
\put(862,25){\makebox(0,0){\large{$l$}}}
\put(350.0,250.0){\rule[-0.200pt]{0.400pt}{361.350pt}}
\put(533.0,780.0){\rule[-0.200pt]{0.400pt}{43.362pt}}
\put(523.0,780.0){\rule[-0.200pt]{4.818pt}{0.400pt}}
\put(523.0,960.0){\rule[-0.200pt]{4.818pt}{0.400pt}}
\put(613.0,892.0){\rule[-0.200pt]{0.400pt}{27.944pt}}
\put(603.0,892.0){\rule[-0.200pt]{4.818pt}{0.400pt}}
\put(603.0,1008.0){\rule[-0.200pt]{4.818pt}{0.400pt}}
\put(721.0,747.0){\rule[-0.200pt]{0.400pt}{38.785pt}}
\put(711.0,747.0){\rule[-0.200pt]{4.818pt}{0.400pt}}
\put(711.0,908.0){\rule[-0.200pt]{4.818pt}{0.400pt}}
\put(882.0,615.0){\rule[-0.200pt]{0.400pt}{29.872pt}}
\put(872.0,615.0){\rule[-0.200pt]{4.818pt}{0.400pt}}
\put(872.0,739.0){\rule[-0.200pt]{4.818pt}{0.400pt}}
\put(1097.0,690.0){\rule[-0.200pt]{0.400pt}{62.393pt}}
\put(1087.0,690.0){\rule[-0.200pt]{4.818pt}{0.400pt}}
\put(1087.0,949.0){\rule[-0.200pt]{4.818pt}{0.400pt}}
\put(506.0,870.0){\rule[-0.200pt]{13.009pt}{0.400pt}}
\put(506.0,860.0){\rule[-0.200pt]{0.400pt}{4.818pt}}
\put(560.0,860.0){\rule[-0.200pt]{0.400pt}{4.818pt}}
\put(560.0,950.0){\rule[-0.200pt]{25.776pt}{0.400pt}}
\put(560.0,940.0){\rule[-0.200pt]{0.400pt}{4.818pt}}
\put(667.0,940.0){\rule[-0.200pt]{0.400pt}{4.818pt}}
\put(667.0,828.0){\rule[-0.200pt]{26.017pt}{0.400pt}}
\put(667.0,818.0){\rule[-0.200pt]{0.400pt}{4.818pt}}
\put(775.0,818.0){\rule[-0.200pt]{0.400pt}{4.818pt}}
\put(775.0,677.0){\rule[-0.200pt]{51.793pt}{0.400pt}}
\put(775.0,667.0){\rule[-0.200pt]{0.400pt}{4.818pt}}
\put(990.0,667.0){\rule[-0.200pt]{0.400pt}{4.818pt}}
\put(990.0,819.0){\rule[-0.200pt]{51.793pt}{0.400pt}}
\put(990.0,809.0){\rule[-0.200pt]{0.400pt}{4.818pt}}
\put(533,870){\circle*{24}}
\put(613,950){\circle*{24}}
\put(721,828){\circle*{24}}
\put(882,677){\circle*{24}}
\put(1097,819){\circle*{24}}
\put(1205.0,809.0){\rule[-0.200pt]{0.400pt}{4.818pt}}
\put(544.0,427.0){\rule[-0.200pt]{0.400pt}{112.741pt}}
\put(534.0,427.0){\rule[-0.200pt]{4.818pt}{0.400pt}}
\put(534.0,895.0){\rule[-0.200pt]{4.818pt}{0.400pt}}
\put(624.0,1290.0){\rule[-0.200pt]{0.400pt}{66.729pt}}
\put(614.0,1290.0){\rule[-0.200pt]{4.818pt}{0.400pt}}
\put(614.0,1567.0){\rule[-0.200pt]{4.818pt}{0.400pt}}
\put(732.0,946.0){\rule[-0.200pt]{0.400pt}{87.447pt}}
\put(722.0,946.0){\rule[-0.200pt]{4.818pt}{0.400pt}}
\put(722.0,1309.0){\rule[-0.200pt]{4.818pt}{0.400pt}}
\put(893.0,696.0){\rule[-0.200pt]{0.400pt}{85.279pt}}
\put(883.0,696.0){\rule[-0.200pt]{4.818pt}{0.400pt}}
\put(883.0,1050.0){\rule[-0.200pt]{4.818pt}{0.400pt}}
\put(517.0,661.0){\rule[-0.200pt]{12.768pt}{0.400pt}}
\put(517.0,651.0){\rule[-0.200pt]{0.400pt}{4.818pt}}
\put(570.0,651.0){\rule[-0.200pt]{0.400pt}{4.818pt}}
\put(570.0,1429.0){\rule[-0.200pt]{26.017pt}{0.400pt}}
\put(570.0,1419.0){\rule[-0.200pt]{0.400pt}{4.818pt}}
\put(678.0,1419.0){\rule[-0.200pt]{0.400pt}{4.818pt}}
\put(678.0,1127.0){\rule[-0.200pt]{25.776pt}{0.400pt}}
\put(678.0,1117.0){\rule[-0.200pt]{0.400pt}{4.818pt}}
\put(785.0,1117.0){\rule[-0.200pt]{0.400pt}{4.818pt}}
\put(785.0,873.0){\rule[-0.200pt]{51.793pt}{0.400pt}}
\put(785.0,863.0){\rule[-0.200pt]{0.400pt}{4.818pt}}
\put(544,661){\circle{24}}
\put(624,1429){\circle{24}}
\put(732,1127){\circle{24}}
\put(893,873){\circle{24}}
\put(1000.0,863.0){\rule[-0.200pt]{0.400pt}{4.818pt}}
\put(350,679){\usebox{\plotpoint}}
\put(350.00,679.00){\usebox{\plotpoint}}
\put(370.76,679.00){\usebox{\plotpoint}}
\put(391.51,679.00){\usebox{\plotpoint}}
\put(412.27,679.00){\usebox{\plotpoint}}
\put(433.02,679.00){\usebox{\plotpoint}}
\put(453.78,679.00){\usebox{\plotpoint}}
\put(474.53,679.00){\usebox{\plotpoint}}
\put(495.29,679.00){\usebox{\plotpoint}}
\put(516.04,679.00){\usebox{\plotpoint}}
\put(536.80,679.00){\usebox{\plotpoint}}
\put(557.55,679.00){\usebox{\plotpoint}}
\put(578.31,679.00){\usebox{\plotpoint}}
\put(599.07,679.00){\usebox{\plotpoint}}
\put(619.82,679.00){\usebox{\plotpoint}}
\put(640.58,679.00){\usebox{\plotpoint}}
\put(661.33,679.00){\usebox{\plotpoint}}
\put(682.09,679.00){\usebox{\plotpoint}}
\put(702.84,679.00){\usebox{\plotpoint}}
\put(723.60,679.00){\usebox{\plotpoint}}
\put(744.35,679.00){\usebox{\plotpoint}}
\put(765.11,679.00){\usebox{\plotpoint}}
\put(785.87,679.00){\usebox{\plotpoint}}
\put(806.62,679.00){\usebox{\plotpoint}}
\put(827.38,679.00){\usebox{\plotpoint}}
\put(848.13,679.00){\usebox{\plotpoint}}
\put(868.89,679.00){\usebox{\plotpoint}}
\put(889.64,679.00){\usebox{\plotpoint}}
\put(910.40,679.00){\usebox{\plotpoint}}
\put(931.15,679.00){\usebox{\plotpoint}}
\put(951.91,679.00){\usebox{\plotpoint}}
\put(972.66,679.00){\usebox{\plotpoint}}
\put(993.42,679.00){\usebox{\plotpoint}}
\put(1014.18,679.00){\usebox{\plotpoint}}
\put(1034.93,679.00){\usebox{\plotpoint}}
\put(1055.69,679.00){\usebox{\plotpoint}}
\put(1076.44,679.00){\usebox{\plotpoint}}
\put(1097.20,679.00){\usebox{\plotpoint}}
\put(1117.95,679.00){\usebox{\plotpoint}}
\put(1138.71,679.00){\usebox{\plotpoint}}
\put(1159.46,679.00){\usebox{\plotpoint}}
\put(1180.22,679.00){\usebox{\plotpoint}}
\put(1200.98,679.00){\usebox{\plotpoint}}
\put(1221.73,679.00){\usebox{\plotpoint}}
\put(1242.49,679.00){\usebox{\plotpoint}}
\put(1263.24,679.00){\usebox{\plotpoint}}
\put(1284.00,679.00){\usebox{\plotpoint}}
\put(1304.75,679.00){\usebox{\plotpoint}}
\put(1325.51,679.00){\usebox{\plotpoint}}
\put(1346.26,679.00){\usebox{\plotpoint}}
\put(1367.02,679.00){\usebox{\plotpoint}}
\put(1387.77,679.00){\usebox{\plotpoint}}
\put(1408.53,679.00){\usebox{\plotpoint}}
\put(1425,679){\usebox{\plotpoint}}
\end{picture}

\end	{center}
\vskip 0.15in
\caption{The effective coefficient of the $1/l$ universal string
correction, calculated from eqn(\ref{eqn_ceff}), for the range 
of lengths indicated.
For $k=1$, $\bullet$, and $k=2$, $\circ$, strings in SU(6).}
\label{fig_ceff}
\end 	{figure}

We now attempt to fit the flux loop mass with a leading linear 
term and a $O(1/l)$ string correction as in eqn(\ref{eqn_corr}).
Since these are supposed to be the leading terms at large $l$,
it is not surprising that we do not obtain an acceptable fit
over the whole range of $l$. We therefore drop lower $l$ values
from the fit and find that we first obtain an acceptable $\chi^2$
per degree of freedom, $\chi^2_{df}$,
for $l\geq 10$, with a slightly better fit for $l\geq 12$:
\begin{equation}
aE_0(l)
= 
a^2\sigma l - \frac{c\pi}{3l} 
=
\begin{cases} 
0.06383(38)l - {1.16(8)\pi}/{3l}
&:\quad l \geq 10 \ \ , \chi^2_{df} = 0.85 \\
0.06356(47)l - {1.09(10)\pi}/{3l}
&:\quad l \geq 12 \ \ , \chi^2_{df} = 0.70
\end{cases}
\label{eqn_fitl2k1n6} 
\end{equation}
Defining an effective value, $c_{eff}$, of the string coefficient
obtained between neighbouring values of $l$
\begin{equation}
c_{eff}\frac{\pi}{3}
\Bigl\{ \frac{1}{l^2_1} -  \frac{1}{l^2_2}\Bigr\}
= 
\frac{aE_0(l_2)}{l_2}-\frac{aE_0(l_1)}{l_1},
\label{eqn_ceff} 
\end{equation}
we show in Fig.~\ref{fig_ceff} how it varies with increasing $l$. 
All this provides significant evidence that the effective string
theory for long flux tubes is in the universality class of
a simple bosonic string theory, $c=1$. 

There will of course be higher order corrections in $1/l$
and, under certain natural assumptions, the next correction
will be down by $1/l^2$
\cite{luscher-string04}.
If we fix the coefficient of the $1/l$ term to the bosonic
string value $\pi/3$, then we find that we can obtain an 
acceptable fit to our whole range of $l$:
\begin{equation}
aE_0(l)
= 
0.06358(20) l - \frac{\pi}{3l} - \frac{19.4(3.2)}{l^3} 
\quad : \quad l \geq 7 \ \ , \chi^2_{df} = 0.9
\label{eqn_fitl4k1n6} 
\end{equation}
This is remarkable since, as pointed out earlier, there is a 
minimum length for a periodic flux loop, which is 
$l_{min} = 1/aT_c \simeq 6.66$ in the present calculation. 
Thus the fit in eqn(\ref{eqn_fitl4k1n6}) essentially works all 
the way down to the shortest possible strings. (If we fit the
$O(1/l)$ term as well, then its coefficient comes to 0.94(16),
with a slightly worse value of $\chi^2_{df}$.) Note that
the natural dimensionless expansion parameter is $1/\sigma l^2$
and so we would expect the $1/l^3$ term in eqn(\ref{eqn_fitl4k1n6})
to have a coefficient that is larger by a factor 
$O(1/\sigma)\sim 16$ than that of the leading $1/l$ correction.
This is indeed consistent with what we find. Note also that
the $O(1/l^3)$ coefficient is about twice as large as the Nambu-Goto  
coefficient in eqn(\ref{eqn_NGE}).

In Fig.\ref{fig_K1} we also plot the prediction of the Nambu-Goto 
string theory for $E_0(l)$, as in eqn(\ref{eqn_NG}) with $n=0$. 
This fit has only one free parameter, $\sigma$, and visually
follows the calculated values quite closely although, with
$\chi^2_{df} \simeq 3.0$, it is clearly not an acceptable fit. 
However, for the comparable SU(4) calculation (discussed below)
we find  $\chi^2_{df} \simeq 7.5$. This improvement in goodness
of fit with increasing $N$ does leave open the
intriguing possibility that at $N = \infty$ the effective string
theory might be governed by the simple Nambu-Goto action.

We also plot the prediction from eqn(\ref{eqn_NGE})
for the first excited state, which now has
no free parameter at all. We see that this tracks the
calculated values quite well, except at the very smallest
$l$. It is interesting to
note that the gap between the energy of the
lightest and first excited string state is large, 
and for most values of $l$ larger than the mass of the 
lightest glueball, which has a mass $am_G\simeq 0.74$.
For such $l$ the excited string can presumably
decay into the lightest string and a glueball. This should
show up in the correlator through an effective energy that
gradually decreases with increasing $t$. The fact that we see no
sign of this is presumably due to the large-$N$ 
suppression of decays. For SU(4), where decays are not
so suppressed,  we do in fact find it difficult to extract 
excited string energies for intermediate $l$, and it is possible
that this is part of the reason.

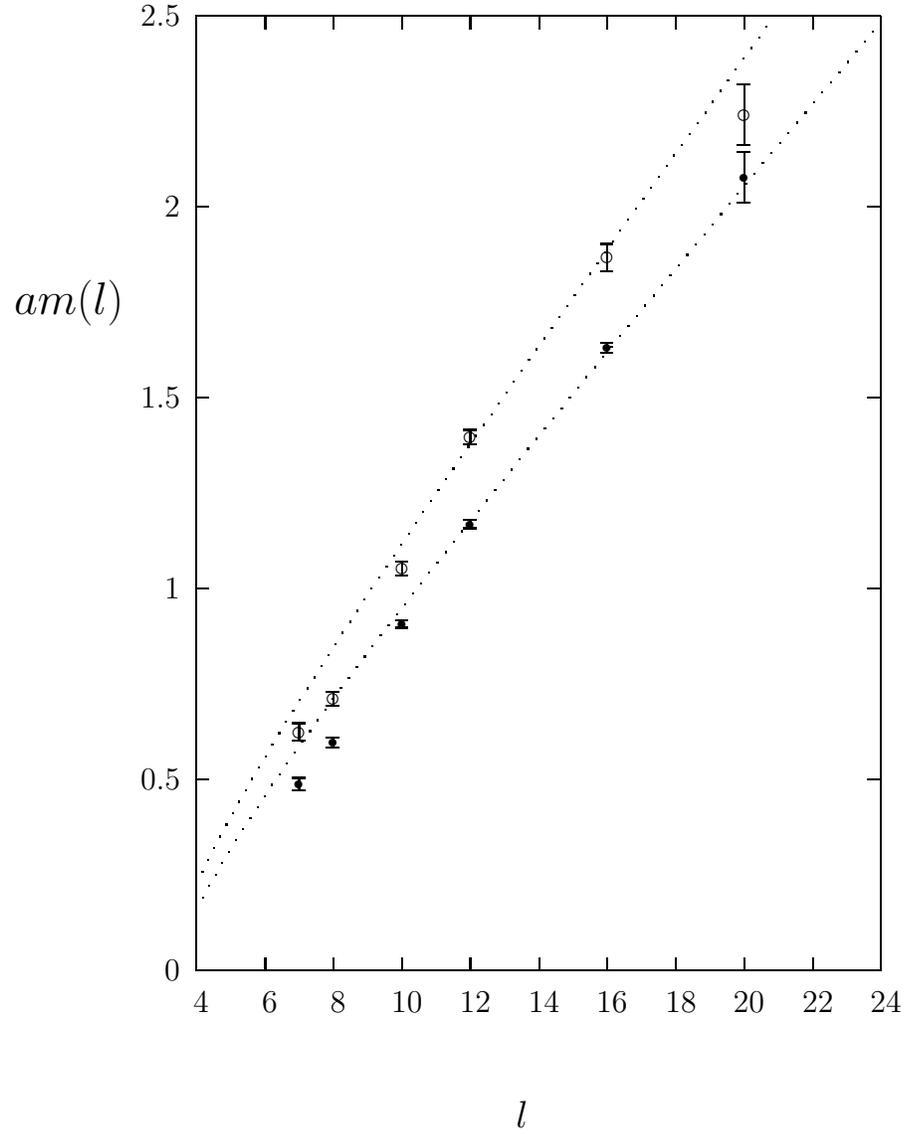
\begin	{figure}[p]
\begin	{center}
\leavevmode
\setlength{\unitlength}{0.240900pt}
\ifx\plotpoint\undefined\newsavebox{\plotpoint}\fi
\sbox{\plotpoint}{\rule[-0.200pt]{0.400pt}{0.400pt}}%
\begin{picture}(1500,1800)(0,0)
\font\gnuplot=cmr10 at 12pt
\gnuplot
\sbox{\plotpoint}{\rule[-0.200pt]{0.400pt}{0.400pt}}%
\put(350.0,250.0){\rule[-0.200pt]{4.818pt}{0.400pt}}
\put(325,250){\makebox(0,0)[r]{\ \ {$0$}}}
\put(1405.0,250.0){\rule[-0.200pt]{4.818pt}{0.400pt}}
\put(350.0,550.0){\rule[-0.200pt]{4.818pt}{0.400pt}}
\put(325,550){\makebox(0,0)[r]{\ \ {$0.5$}}}
\put(1405.0,550.0){\rule[-0.200pt]{4.818pt}{0.400pt}}
\put(350.0,850.0){\rule[-0.200pt]{4.818pt}{0.400pt}}
\put(325,850){\makebox(0,0)[r]{\ \ {$1$}}}
\put(1405.0,850.0){\rule[-0.200pt]{4.818pt}{0.400pt}}
\put(350.0,1150.0){\rule[-0.200pt]{4.818pt}{0.400pt}}
\put(325,1150){\makebox(0,0)[r]{\ \ {$1.5$}}}
\put(1405.0,1150.0){\rule[-0.200pt]{4.818pt}{0.400pt}}
\put(350.0,1450.0){\rule[-0.200pt]{4.818pt}{0.400pt}}
\put(325,1450){\makebox(0,0)[r]{\ \ {$2$}}}
\put(1405.0,1450.0){\rule[-0.200pt]{4.818pt}{0.400pt}}
\put(350.0,1750.0){\rule[-0.200pt]{4.818pt}{0.400pt}}
\put(325,1750){\makebox(0,0)[r]{\ \ {$2.5$}}}
\put(1405.0,1750.0){\rule[-0.200pt]{4.818pt}{0.400pt}}
\put(350.0,250.0){\rule[-0.200pt]{0.400pt}{4.818pt}}
\put(350,200){\makebox(0,0){\ {$4$}}}
\put(350.0,1730.0){\rule[-0.200pt]{0.400pt}{4.818pt}}
\put(458.0,250.0){\rule[-0.200pt]{0.400pt}{4.818pt}}
\put(458,200){\makebox(0,0){\ {$6$}}}
\put(458.0,1730.0){\rule[-0.200pt]{0.400pt}{4.818pt}}
\put(565.0,250.0){\rule[-0.200pt]{0.400pt}{4.818pt}}
\put(565,200){\makebox(0,0){\ {$8$}}}
\put(565.0,1730.0){\rule[-0.200pt]{0.400pt}{4.818pt}}
\put(673.0,250.0){\rule[-0.200pt]{0.400pt}{4.818pt}}
\put(673,200){\makebox(0,0){\ {$10$}}}
\put(673.0,1730.0){\rule[-0.200pt]{0.400pt}{4.818pt}}
\put(780.0,250.0){\rule[-0.200pt]{0.400pt}{4.818pt}}
\put(780,200){\makebox(0,0){\ {$12$}}}
\put(780.0,1730.0){\rule[-0.200pt]{0.400pt}{4.818pt}}
\put(888.0,250.0){\rule[-0.200pt]{0.400pt}{4.818pt}}
\put(888,200){\makebox(0,0){\ {$14$}}}
\put(888.0,1730.0){\rule[-0.200pt]{0.400pt}{4.818pt}}
\put(995.0,250.0){\rule[-0.200pt]{0.400pt}{4.818pt}}
\put(995,200){\makebox(0,0){\ {$16$}}}
\put(995.0,1730.0){\rule[-0.200pt]{0.400pt}{4.818pt}}
\put(1103.0,250.0){\rule[-0.200pt]{0.400pt}{4.818pt}}
\put(1103,200){\makebox(0,0){\ {$18$}}}
\put(1103.0,1730.0){\rule[-0.200pt]{0.400pt}{4.818pt}}
\put(1210.0,250.0){\rule[-0.200pt]{0.400pt}{4.818pt}}
\put(1210,200){\makebox(0,0){\ {$20$}}}
\put(1210.0,1730.0){\rule[-0.200pt]{0.400pt}{4.818pt}}
\put(1318.0,250.0){\rule[-0.200pt]{0.400pt}{4.818pt}}
\put(1318,200){\makebox(0,0){\ {$22$}}}
\put(1318.0,1730.0){\rule[-0.200pt]{0.400pt}{4.818pt}}
\put(1425.0,250.0){\rule[-0.200pt]{0.400pt}{4.818pt}}
\put(1425,200){\makebox(0,0){\ {$24$}}}
\put(1425.0,1730.0){\rule[-0.200pt]{0.400pt}{4.818pt}}
\put(350.0,250.0){\rule[-0.200pt]{258.967pt}{0.400pt}}
\put(1425.0,250.0){\rule[-0.200pt]{0.400pt}{361.350pt}}
\put(350.0,1750.0){\rule[-0.200pt]{258.967pt}{0.400pt}}
\put(150,1300){\makebox(0,0){\Large{$am(l)$}}}
\put(862,25){\makebox(0,0){\large{$l$}}}
\put(350.0,250.0){\rule[-0.200pt]{0.400pt}{361.350pt}}
\put(511.0,533.0){\rule[-0.200pt]{0.400pt}{4.577pt}}
\put(501.0,533.0){\rule[-0.200pt]{4.818pt}{0.400pt}}
\put(501.0,552.0){\rule[-0.200pt]{4.818pt}{0.400pt}}
\put(565.0,600.0){\rule[-0.200pt]{0.400pt}{3.613pt}}
\put(555.0,600.0){\rule[-0.200pt]{4.818pt}{0.400pt}}
\put(555.0,615.0){\rule[-0.200pt]{4.818pt}{0.400pt}}
\put(673.0,788.0){\rule[-0.200pt]{0.400pt}{2.891pt}}
\put(663.0,788.0){\rule[-0.200pt]{4.818pt}{0.400pt}}
\put(663.0,800.0){\rule[-0.200pt]{4.818pt}{0.400pt}}
\put(780.0,944.0){\rule[-0.200pt]{0.400pt}{3.132pt}}
\put(770.0,944.0){\rule[-0.200pt]{4.818pt}{0.400pt}}
\put(770.0,957.0){\rule[-0.200pt]{4.818pt}{0.400pt}}
\put(995.0,1220.0){\rule[-0.200pt]{0.400pt}{3.854pt}}
\put(985.0,1220.0){\rule[-0.200pt]{4.818pt}{0.400pt}}
\put(985.0,1236.0){\rule[-0.200pt]{4.818pt}{0.400pt}}
\put(1210.0,1456.0){\rule[-0.200pt]{0.400pt}{19.272pt}}
\put(1200.0,1456.0){\rule[-0.200pt]{4.818pt}{0.400pt}}
\put(511,543){\circle*{12}}
\put(565,608){\circle*{12}}
\put(673,794){\circle*{12}}
\put(780,950){\circle*{12}}
\put(995,1228){\circle*{12}}
\put(1210,1496){\circle*{12}}
\put(1200.0,1536.0){\rule[-0.200pt]{4.818pt}{0.400pt}}
\put(511.0,611.0){\rule[-0.200pt]{0.400pt}{6.504pt}}
\put(501.0,611.0){\rule[-0.200pt]{4.818pt}{0.400pt}}
\put(501.0,638.0){\rule[-0.200pt]{4.818pt}{0.400pt}}
\put(565.0,665.0){\rule[-0.200pt]{0.400pt}{5.300pt}}
\put(555.0,665.0){\rule[-0.200pt]{4.818pt}{0.400pt}}
\put(555.0,687.0){\rule[-0.200pt]{4.818pt}{0.400pt}}
\put(673.0,870.0){\rule[-0.200pt]{0.400pt}{5.300pt}}
\put(663.0,870.0){\rule[-0.200pt]{4.818pt}{0.400pt}}
\put(663.0,892.0){\rule[-0.200pt]{4.818pt}{0.400pt}}
\put(780.0,1076.0){\rule[-0.200pt]{0.400pt}{5.541pt}}
\put(770.0,1076.0){\rule[-0.200pt]{4.818pt}{0.400pt}}
\put(770.0,1099.0){\rule[-0.200pt]{4.818pt}{0.400pt}}
\put(995.0,1348.0){\rule[-0.200pt]{0.400pt}{10.359pt}}
\put(985.0,1348.0){\rule[-0.200pt]{4.818pt}{0.400pt}}
\put(985.0,1391.0){\rule[-0.200pt]{4.818pt}{0.400pt}}
\put(1210.0,1546.0){\rule[-0.200pt]{0.400pt}{23.126pt}}
\put(1200.0,1546.0){\rule[-0.200pt]{4.818pt}{0.400pt}}
\put(511,624){\circle{18}}
\put(565,676){\circle{18}}
\put(673,881){\circle{18}}
\put(780,1088){\circle{18}}
\put(995,1370){\circle{18}}
\put(1210,1594){\circle{18}}
\put(1200.0,1642.0){\rule[-0.200pt]{4.818pt}{0.400pt}}
\put(350,346){\usebox{\plotpoint}}
\multiput(350,346)(10.002,18.186){2}{\usebox{\plotpoint}}
\put(370.00,382.37){\usebox{\plotpoint}}
\put(380.33,400.38){\usebox{\plotpoint}}
\put(390.49,418.48){\usebox{\plotpoint}}
\put(401.12,436.29){\usebox{\plotpoint}}
\put(411.95,454.01){\usebox{\plotpoint}}
\put(423.10,471.51){\usebox{\plotpoint}}
\put(434.37,488.94){\usebox{\plotpoint}}
\put(445.65,506.36){\usebox{\plotpoint}}
\put(457.30,523.53){\usebox{\plotpoint}}
\put(468.41,541.06){\usebox{\plotpoint}}
\put(480.13,558.19){\usebox{\plotpoint}}
\put(491.93,575.27){\usebox{\plotpoint}}
\put(504.11,592.07){\usebox{\plotpoint}}
\put(516.00,609.08){\usebox{\plotpoint}}
\put(528.27,625.82){\usebox{\plotpoint}}
\put(540.20,642.80){\usebox{\plotpoint}}
\put(552.16,659.76){\usebox{\plotpoint}}
\put(564.43,676.50){\usebox{\plotpoint}}
\put(576.70,693.23){\usebox{\plotpoint}}
\put(589.45,709.61){\usebox{\plotpoint}}
\put(601.80,726.29){\usebox{\plotpoint}}
\put(614.41,742.77){\usebox{\plotpoint}}
\put(626.57,759.59){\usebox{\plotpoint}}
\put(639.15,776.10){\usebox{\plotpoint}}
\put(651.97,792.42){\usebox{\plotpoint}}
\put(664.80,808.74){\usebox{\plotpoint}}
\put(677.62,825.06){\usebox{\plotpoint}}
\put(690.24,841.53){\usebox{\plotpoint}}
\put(702.64,858.17){\usebox{\plotpoint}}
\put(715.46,874.49){\usebox{\plotpoint}}
\put(728.28,890.81){\usebox{\plotpoint}}
\put(741.58,906.74){\usebox{\plotpoint}}
\put(754.41,923.06){\usebox{\plotpoint}}
\put(766.98,939.57){\usebox{\plotpoint}}
\put(779.72,955.94){\usebox{\plotpoint}}
\put(792.73,972.11){\usebox{\plotpoint}}
\put(806.03,988.04){\usebox{\plotpoint}}
\put(818.85,1004.36){\usebox{\plotpoint}}
\put(831.84,1020.54){\usebox{\plotpoint}}
\put(844.62,1036.87){\usebox{\plotpoint}}
\put(857.54,1053.10){\usebox{\plotpoint}}
\put(870.47,1069.33){\usebox{\plotpoint}}
\put(883.86,1085.19){\usebox{\plotpoint}}
\put(897.08,1101.19){\usebox{\plotpoint}}
\put(910.17,1117.29){\usebox{\plotpoint}}
\put(923.58,1133.13){\usebox{\plotpoint}}
\put(935.88,1149.84){\usebox{\plotpoint}}
\put(949.28,1165.69){\usebox{\plotpoint}}
\put(962.48,1181.70){\usebox{\plotpoint}}
\put(975.59,1197.79){\usebox{\plotpoint}}
\put(989.00,1213.63){\usebox{\plotpoint}}
\put(1001.91,1229.88){\usebox{\plotpoint}}
\put(1014.71,1246.21){\usebox{\plotpoint}}
\put(1028.12,1262.05){\usebox{\plotpoint}}
\put(1041.53,1277.90){\usebox{\plotpoint}}
\put(1054.50,1294.09){\usebox{\plotpoint}}
\put(1067.84,1309.99){\usebox{\plotpoint}}
\put(1081.07,1325.99){\usebox{\plotpoint}}
\put(1094.06,1342.16){\usebox{\plotpoint}}
\put(1107.10,1358.31){\usebox{\plotpoint}}
\put(1120.37,1374.26){\usebox{\plotpoint}}
\put(1133.78,1390.10){\usebox{\plotpoint}}
\put(1147.19,1405.95){\usebox{\plotpoint}}
\put(1160.22,1422.09){\usebox{\plotpoint}}
\put(1173.41,1438.12){\usebox{\plotpoint}}
\put(1186.81,1453.96){\usebox{\plotpoint}}
\put(1200.08,1469.92){\usebox{\plotpoint}}
\put(1213.13,1486.06){\usebox{\plotpoint}}
\put(1226.53,1501.90){\usebox{\plotpoint}}
\put(1239.38,1518.20){\usebox{\plotpoint}}
\put(1252.75,1534.07){\usebox{\plotpoint}}
\put(1266.16,1549.92){\usebox{\plotpoint}}
\put(1279.57,1565.76){\usebox{\plotpoint}}
\put(1292.97,1581.61){\usebox{\plotpoint}}
\put(1306.36,1597.47){\usebox{\plotpoint}}
\put(1319.19,1613.77){\usebox{\plotpoint}}
\put(1332.60,1629.62){\usebox{\plotpoint}}
\put(1346.01,1645.46){\usebox{\plotpoint}}
\put(1359.41,1661.31){\usebox{\plotpoint}}
\put(1372.82,1677.15){\usebox{\plotpoint}}
\put(1385.99,1693.19){\usebox{\plotpoint}}
\put(1399.04,1709.32){\usebox{\plotpoint}}
\put(1412.45,1725.16){\usebox{\plotpoint}}
\put(1425,1740){\usebox{\plotpoint}}
\put(350,386){\usebox{\plotpoint}}
\multiput(350,386)(9.282,18.564){2}{\usebox{\plotpoint}}
\put(368.56,423.13){\usebox{\plotpoint}}
\put(378.07,441.58){\usebox{\plotpoint}}
\put(387.35,460.14){\usebox{\plotpoint}}
\put(396.67,478.68){\usebox{\plotpoint}}
\put(406.78,496.80){\usebox{\plotpoint}}
\put(417.10,514.81){\usebox{\plotpoint}}
\put(427.14,532.97){\usebox{\plotpoint}}
\put(437.56,550.92){\usebox{\plotpoint}}
\multiput(448,568)(10.399,17.962){2}{\usebox{\plotpoint}}
\put(468.86,604.74){\usebox{\plotpoint}}
\put(479.67,622.46){\usebox{\plotpoint}}
\put(490.49,640.17){\usebox{\plotpoint}}
\put(501.75,657.61){\usebox{\plotpoint}}
\put(512.58,675.31){\usebox{\plotpoint}}
\put(523.84,692.75){\usebox{\plotpoint}}
\put(535.11,710.18){\usebox{\plotpoint}}
\put(545.67,728.04){\usebox{\plotpoint}}
\put(556.95,745.47){\usebox{\plotpoint}}
\put(568.22,762.89){\usebox{\plotpoint}}
\put(579.56,780.28){\usebox{\plotpoint}}
\put(591.23,797.44){\usebox{\plotpoint}}
\put(602.61,814.80){\usebox{\plotpoint}}
\put(614.02,832.13){\usebox{\plotpoint}}
\put(624.95,849.75){\usebox{\plotpoint}}
\put(636.71,866.85){\usebox{\plotpoint}}
\put(648.47,883.96){\usebox{\plotpoint}}
\put(659.97,901.23){\usebox{\plotpoint}}
\put(671.52,918.48){\usebox{\plotpoint}}
\put(683.28,935.58){\usebox{\plotpoint}}
\put(694.52,953.03){\usebox{\plotpoint}}
\put(706.10,970.24){\usebox{\plotpoint}}
\put(718.29,987.04){\usebox{\plotpoint}}
\put(730.08,1004.12){\usebox{\plotpoint}}
\put(741.84,1021.22){\usebox{\plotpoint}}
\put(753.60,1038.33){\usebox{\plotpoint}}
\put(765.31,1055.46){\usebox{\plotpoint}}
\put(776.90,1072.68){\usebox{\plotpoint}}
\put(788.66,1089.78){\usebox{\plotpoint}}
\put(800.66,1106.72){\usebox{\plotpoint}}
\put(812.64,1123.66){\usebox{\plotpoint}}
\put(824.40,1140.76){\usebox{\plotpoint}}
\put(836.52,1157.61){\usebox{\plotpoint}}
\put(847.77,1175.04){\usebox{\plotpoint}}
\put(859.91,1191.87){\usebox{\plotpoint}}
\put(871.70,1208.95){\usebox{\plotpoint}}
\put(883.97,1225.69){\usebox{\plotpoint}}
\put(896.11,1242.52){\usebox{\plotpoint}}
\put(908.04,1259.51){\usebox{\plotpoint}}
\put(920.09,1276.40){\usebox{\plotpoint}}
\put(931.73,1293.59){\usebox{\plotpoint}}
\put(943.72,1310.52){\usebox{\plotpoint}}
\put(955.61,1327.53){\usebox{\plotpoint}}
\put(967.78,1344.34){\usebox{\plotpoint}}
\put(980.05,1361.08){\usebox{\plotpoint}}
\put(991.85,1378.16){\usebox{\plotpoint}}
\put(1003.99,1394.99){\usebox{\plotpoint}}
\put(1015.58,1412.21){\usebox{\plotpoint}}
\put(1027.53,1429.17){\usebox{\plotpoint}}
\put(1039.80,1445.91){\usebox{\plotpoint}}
\put(1052.08,1462.65){\usebox{\plotpoint}}
\put(1064.00,1479.64){\usebox{\plotpoint}}
\put(1076.14,1496.47){\usebox{\plotpoint}}
\put(1087.77,1513.66){\usebox{\plotpoint}}
\put(1100.03,1530.40){\usebox{\plotpoint}}
\put(1112.21,1547.21){\usebox{\plotpoint}}
\put(1124.10,1564.22){\usebox{\plotpoint}}
\put(1136.37,1580.96){\usebox{\plotpoint}}
\put(1148.64,1597.70){\usebox{\plotpoint}}
\put(1160.49,1614.73){\usebox{\plotpoint}}
\put(1172.53,1631.63){\usebox{\plotpoint}}
\put(1184.39,1648.66){\usebox{\plotpoint}}
\put(1196.60,1665.45){\usebox{\plotpoint}}
\put(1208.87,1682.19){\usebox{\plotpoint}}
\put(1221.14,1698.92){\usebox{\plotpoint}}
\put(1233.21,1715.81){\usebox{\plotpoint}}
\put(1245.03,1732.86){\usebox{\plotpoint}}
\put(1257.55,1749.42){\usebox{\plotpoint}}
\put(1258,1750){\usebox{\plotpoint}}
\end{picture}

\end	{center}
\vskip 0.15in
\caption{The masses of the lightest $k=2$, $\bullet$, and  $k=3$, 
$\circ$, flux loops that wind
around a spatial torus of length $l$ in the SU(6) 
calculation at $\beta=25.05$. The dotted lines are the
best fits with a bosonic string correction,
as in eqn(\ref{eqn_corr}) with $c=1$.}
\label{fig_K23}
\end 	{figure}

In Fig.\ref{fig_K23} we plot the masses of the lightest
$k=2$ and $k=3$ strings. These are heavier than the $k=1$ string
and thus possess
larger errors. Nonetheless the approximate linear growth with 
$l$ is evident. We plot fits with the bosonic string correction
and we list the best coefficients of a $1/l$ correction
in Table~\ref{table_su6fit}. The $k=3$ calculation is not accurate
enough to be informative on this issue, but for the 
$k=2$ string we have some statistically weak
evidence that for $l\geq 12a \simeq 3/\surd\sigma$ the 
dominant correction is $O(1/l)$, with a coefficient that is
consistent with that of the simple bosonic string. To investigate
this in more detail we analyse the $k=2$ string masses using 
eqn(\ref{eqn_ceff}), and plot the result in Fig.~\ref{fig_ceff}.
Comparing the values of $c_{eff}$ from the $k=1$ and $k=2$ strings, 
we see that the $k=2$ values are much larger for shorter strings, 
and, while consistent with approaching the bosonic string value,
probably does so for somewhat larger values of $l$. 
This behaviour is, in fact, not unexpected if we think of 
the $k=2$ string as two loosely bound $k=1$ strings. If it
acts as two incoherent, unbound strings it receives twice
the string correction that it would receive if it behaved as
a coherent, single string. Since the correction is negative
and grows with decreasing $l$, it is clear that there should 
be some length $l^\prime$ such that for $l<l^\prime$ 
it will be energetically favourable for the $k$-string 
to be $k$ incoherent $k=1$ strings rather than a single
coherent $k$-string. Moreover since
$\sigma_k \to k \sigma$ as $N\to \infty$, it is also
clear that $l^\prime \to \infty$ as $N\to \infty$
\cite{blmtuw-glue04}. 
More generally we can think of the L\"uscher
correction as inducing a repulsive interaction between
the two strings, which leads to a value of $c_{eff}$
that smoothly decreases from 2 to 1 at a length scale that
grows with $N$. It is tempting to see Fig.~\ref{fig_ceff}
as displaying such a behaviour.

\begin	{figure}[p]
\begin	{center}
\leavevmode
\setlength{\unitlength}{0.240900pt}
\ifx\plotpoint\undefined\newsavebox{\plotpoint}\fi
\sbox{\plotpoint}{\rule[-0.200pt]{0.400pt}{0.400pt}}%
\begin{picture}(1500,1800)(0,0)
\font\gnuplot=cmr10 at 12pt
\gnuplot
\sbox{\plotpoint}{\rule[-0.200pt]{0.400pt}{0.400pt}}%
\put(350.0,250.0){\rule[-0.200pt]{4.818pt}{0.400pt}}
\put(325,250){\makebox(0,0)[r]{\ \ {$0$}}}
\put(1405.0,250.0){\rule[-0.200pt]{4.818pt}{0.400pt}}
\put(350.0,400.0){\rule[-0.200pt]{4.818pt}{0.400pt}}
\put(325,400){\makebox(0,0)[r]{\ \ {$0.2$}}}
\put(1405.0,400.0){\rule[-0.200pt]{4.818pt}{0.400pt}}
\put(350.0,550.0){\rule[-0.200pt]{4.818pt}{0.400pt}}
\put(325,550){\makebox(0,0)[r]{\ \ {$0.4$}}}
\put(1405.0,550.0){\rule[-0.200pt]{4.818pt}{0.400pt}}
\put(350.0,700.0){\rule[-0.200pt]{4.818pt}{0.400pt}}
\put(325,700){\makebox(0,0)[r]{\ \ {$0.6$}}}
\put(1405.0,700.0){\rule[-0.200pt]{4.818pt}{0.400pt}}
\put(350.0,850.0){\rule[-0.200pt]{4.818pt}{0.400pt}}
\put(325,850){\makebox(0,0)[r]{\ \ {$0.8$}}}
\put(1405.0,850.0){\rule[-0.200pt]{4.818pt}{0.400pt}}
\put(350.0,1000.0){\rule[-0.200pt]{4.818pt}{0.400pt}}
\put(325,1000){\makebox(0,0)[r]{\ \ {$1$}}}
\put(1405.0,1000.0){\rule[-0.200pt]{4.818pt}{0.400pt}}
\put(350.0,1150.0){\rule[-0.200pt]{4.818pt}{0.400pt}}
\put(325,1150){\makebox(0,0)[r]{\ \ {$1.2$}}}
\put(1405.0,1150.0){\rule[-0.200pt]{4.818pt}{0.400pt}}
\put(350.0,1300.0){\rule[-0.200pt]{4.818pt}{0.400pt}}
\put(325,1300){\makebox(0,0)[r]{\ \ {$1.4$}}}
\put(1405.0,1300.0){\rule[-0.200pt]{4.818pt}{0.400pt}}
\put(350.0,1450.0){\rule[-0.200pt]{4.818pt}{0.400pt}}
\put(325,1450){\makebox(0,0)[r]{\ \ {$1.6$}}}
\put(1405.0,1450.0){\rule[-0.200pt]{4.818pt}{0.400pt}}
\put(350.0,1600.0){\rule[-0.200pt]{4.818pt}{0.400pt}}
\put(325,1600){\makebox(0,0)[r]{\ \ {$1.8$}}}
\put(1405.0,1600.0){\rule[-0.200pt]{4.818pt}{0.400pt}}
\put(350.0,1750.0){\rule[-0.200pt]{4.818pt}{0.400pt}}
\put(325,1750){\makebox(0,0)[r]{\ \ {$2$}}}
\put(1405.0,1750.0){\rule[-0.200pt]{4.818pt}{0.400pt}}
\put(350.0,250.0){\rule[-0.200pt]{0.400pt}{4.818pt}}
\put(350,200){\makebox(0,0){\ {$4$}}}
\put(350.0,1730.0){\rule[-0.200pt]{0.400pt}{4.818pt}}
\put(458.0,250.0){\rule[-0.200pt]{0.400pt}{4.818pt}}
\put(458,200){\makebox(0,0){\ {$6$}}}
\put(458.0,1730.0){\rule[-0.200pt]{0.400pt}{4.818pt}}
\put(565.0,250.0){\rule[-0.200pt]{0.400pt}{4.818pt}}
\put(565,200){\makebox(0,0){\ {$8$}}}
\put(565.0,1730.0){\rule[-0.200pt]{0.400pt}{4.818pt}}
\put(673.0,250.0){\rule[-0.200pt]{0.400pt}{4.818pt}}
\put(673,200){\makebox(0,0){\ {$10$}}}
\put(673.0,1730.0){\rule[-0.200pt]{0.400pt}{4.818pt}}
\put(780.0,250.0){\rule[-0.200pt]{0.400pt}{4.818pt}}
\put(780,200){\makebox(0,0){\ {$12$}}}
\put(780.0,1730.0){\rule[-0.200pt]{0.400pt}{4.818pt}}
\put(888.0,250.0){\rule[-0.200pt]{0.400pt}{4.818pt}}
\put(888,200){\makebox(0,0){\ {$14$}}}
\put(888.0,1730.0){\rule[-0.200pt]{0.400pt}{4.818pt}}
\put(995.0,250.0){\rule[-0.200pt]{0.400pt}{4.818pt}}
\put(995,200){\makebox(0,0){\ {$16$}}}
\put(995.0,1730.0){\rule[-0.200pt]{0.400pt}{4.818pt}}
\put(1103.0,250.0){\rule[-0.200pt]{0.400pt}{4.818pt}}
\put(1103,200){\makebox(0,0){\ {$18$}}}
\put(1103.0,1730.0){\rule[-0.200pt]{0.400pt}{4.818pt}}
\put(1210.0,250.0){\rule[-0.200pt]{0.400pt}{4.818pt}}
\put(1210,200){\makebox(0,0){\ {$20$}}}
\put(1210.0,1730.0){\rule[-0.200pt]{0.400pt}{4.818pt}}
\put(1318.0,250.0){\rule[-0.200pt]{0.400pt}{4.818pt}}
\put(1318,200){\makebox(0,0){\ {$22$}}}
\put(1318.0,1730.0){\rule[-0.200pt]{0.400pt}{4.818pt}}
\put(1425.0,250.0){\rule[-0.200pt]{0.400pt}{4.818pt}}
\put(1425,200){\makebox(0,0){\ {$24$}}}
\put(1425.0,1730.0){\rule[-0.200pt]{0.400pt}{4.818pt}}
\put(350.0,250.0){\rule[-0.200pt]{258.967pt}{0.400pt}}
\put(1425.0,250.0){\rule[-0.200pt]{0.400pt}{361.350pt}}
\put(350.0,1750.0){\rule[-0.200pt]{258.967pt}{0.400pt}}
\put(150,1300){\makebox(0,0){\Large{$am(l)$}}}
\put(862,25){\makebox(0,0){\large{$l$}}}
\put(350.0,250.0){\rule[-0.200pt]{0.400pt}{361.350pt}}
\put(511.0,369.0){\rule[-0.200pt]{0.400pt}{2.650pt}}
\put(501.0,369.0){\rule[-0.200pt]{4.818pt}{0.400pt}}
\put(501.0,380.0){\rule[-0.200pt]{4.818pt}{0.400pt}}
\put(565.0,472.0){\rule[-0.200pt]{0.400pt}{1.927pt}}
\put(555.0,472.0){\rule[-0.200pt]{4.818pt}{0.400pt}}
\put(555.0,480.0){\rule[-0.200pt]{4.818pt}{0.400pt}}
\put(673.0,583.0){\rule[-0.200pt]{0.400pt}{2.409pt}}
\put(663.0,583.0){\rule[-0.200pt]{4.818pt}{0.400pt}}
\put(663.0,593.0){\rule[-0.200pt]{4.818pt}{0.400pt}}
\put(780.0,688.0){\rule[-0.200pt]{0.400pt}{3.132pt}}
\put(770.0,688.0){\rule[-0.200pt]{4.818pt}{0.400pt}}
\put(770.0,701.0){\rule[-0.200pt]{4.818pt}{0.400pt}}
\put(995.0,892.0){\rule[-0.200pt]{0.400pt}{1.927pt}}
\put(985.0,892.0){\rule[-0.200pt]{4.818pt}{0.400pt}}
\put(511,374){\circle*{12}}
\put(565,476){\circle*{12}}
\put(673,588){\circle*{12}}
\put(780,695){\circle*{12}}
\put(995,896){\circle*{12}}
\put(985.0,900.0){\rule[-0.200pt]{4.818pt}{0.400pt}}
\put(511.0,1338.0){\rule[-0.200pt]{0.400pt}{18.067pt}}
\put(501.0,1338.0){\rule[-0.200pt]{4.818pt}{0.400pt}}
\put(501.0,1413.0){\rule[-0.200pt]{4.818pt}{0.400pt}}
\put(780.0,1267.0){\rule[-0.200pt]{0.400pt}{6.986pt}}
\put(770.0,1267.0){\rule[-0.200pt]{4.818pt}{0.400pt}}
\put(770.0,1296.0){\rule[-0.200pt]{4.818pt}{0.400pt}}
\put(995.0,1422.0){\rule[-0.200pt]{0.400pt}{2.891pt}}
\put(985.0,1422.0){\rule[-0.200pt]{4.818pt}{0.400pt}}
\put(511,1375){\circle{18}}
\put(780,1281){\circle{18}}
\put(995,1428){\circle{18}}
\put(985.0,1434.0){\rule[-0.200pt]{4.818pt}{0.400pt}}
\put(459,265){\usebox{\plotpoint}}
\multiput(459,265)(3.713,20.421){3}{\usebox{\plotpoint}}
\multiput(469,320)(7.361,19.406){2}{\usebox{\plotpoint}}
\put(487.54,364.08){\usebox{\plotpoint}}
\put(497.27,382.41){\usebox{\plotpoint}}
\put(507.71,400.34){\usebox{\plotpoint}}
\put(519.01,417.74){\usebox{\plotpoint}}
\put(531.07,434.63){\usebox{\plotpoint}}
\put(542.82,451.73){\usebox{\plotpoint}}
\put(555.40,468.23){\usebox{\plotpoint}}
\put(568.78,484.10){\usebox{\plotpoint}}
\put(582.18,499.95){\usebox{\plotpoint}}
\put(595.90,515.52){\usebox{\plotpoint}}
\put(609.92,530.82){\usebox{\plotpoint}}
\put(623.39,546.61){\usebox{\plotpoint}}
\put(637.67,561.67){\usebox{\plotpoint}}
\put(651.93,576.74){\usebox{\plotpoint}}
\put(666.51,591.51){\usebox{\plotpoint}}
\put(681.18,606.18){\usebox{\plotpoint}}
\put(695.43,621.27){\usebox{\plotpoint}}
\put(710.51,635.51){\usebox{\plotpoint}}
\put(725.48,649.89){\usebox{\plotpoint}}
\put(740.35,664.35){\usebox{\plotpoint}}
\put(755.68,678.35){\usebox{\plotpoint}}
\put(770.31,693.04){\usebox{\plotpoint}}
\put(785.40,707.27){\usebox{\plotpoint}}
\put(800.76,721.23){\usebox{\plotpoint}}
\put(816.11,735.19){\usebox{\plotpoint}}
\put(831.47,749.16){\usebox{\plotpoint}}
\put(846.48,763.48){\usebox{\plotpoint}}
\put(862.21,777.01){\usebox{\plotpoint}}
\put(877.56,790.97){\usebox{\plotpoint}}
\put(892.92,804.93){\usebox{\plotpoint}}
\put(908.76,818.33){\usebox{\plotpoint}}
\put(924.12,832.29){\usebox{\plotpoint}}
\put(939.52,846.20){\usebox{\plotpoint}}
\put(955.24,859.75){\usebox{\plotpoint}}
\put(970.72,873.57){\usebox{\plotpoint}}
\put(986.36,887.20){\usebox{\plotpoint}}
\put(1002.39,900.39){\usebox{\plotpoint}}
\put(1017.54,914.54){\usebox{\plotpoint}}
\put(1033.14,928.22){\usebox{\plotpoint}}
\put(1048.98,941.62){\usebox{\plotpoint}}
\put(1064.72,955.14){\usebox{\plotpoint}}
\put(1080.55,968.55){\usebox{\plotpoint}}
\put(1095.91,982.47){\usebox{\plotpoint}}
\put(1111.88,995.71){\usebox{\plotpoint}}
\put(1127.53,1009.34){\usebox{\plotpoint}}
\put(1143.59,1022.48){\usebox{\plotpoint}}
\put(1159.17,1036.17){\usebox{\plotpoint}}
\put(1174.77,1049.81){\usebox{\plotpoint}}
\put(1190.84,1062.96){\usebox{\plotpoint}}
\put(1206.47,1076.61){\usebox{\plotpoint}}
\put(1222.46,1089.83){\usebox{\plotpoint}}
\put(1238.19,1103.37){\usebox{\plotpoint}}
\put(1254.04,1116.76){\usebox{\plotpoint}}
\put(1269.73,1130.33){\usebox{\plotpoint}}
\put(1285.80,1143.47){\usebox{\plotpoint}}
\put(1301.86,1156.61){\usebox{\plotpoint}}
\put(1317.51,1170.24){\usebox{\plotpoint}}
\put(1333.29,1183.72){\usebox{\plotpoint}}
\put(1349.13,1197.11){\usebox{\plotpoint}}
\put(1365.20,1210.25){\usebox{\plotpoint}}
\put(1381.26,1223.40){\usebox{\plotpoint}}
\put(1396.91,1237.02){\usebox{\plotpoint}}
\put(1412.98,1250.16){\usebox{\plotpoint}}
\put(1425,1260){\usebox{\plotpoint}}
\put(350,1134){\usebox{\plotpoint}}
\put(350.00,1134.00){\usebox{\plotpoint}}
\put(370.42,1137.71){\usebox{\plotpoint}}
\put(390.82,1141.56){\usebox{\plotpoint}}
\put(411.23,1145.31){\usebox{\plotpoint}}
\put(431.54,1149.51){\usebox{\plotpoint}}
\put(451.85,1153.70){\usebox{\plotpoint}}
\put(471.94,1158.80){\usebox{\plotpoint}}
\put(492.18,1163.32){\usebox{\plotpoint}}
\put(512.20,1168.78){\usebox{\plotpoint}}
\put(532.23,1174.24){\usebox{\plotpoint}}
\put(552.18,1179.96){\usebox{\plotpoint}}
\put(571.91,1186.34){\usebox{\plotpoint}}
\put(591.86,1192.04){\usebox{\plotpoint}}
\put(611.64,1198.26){\usebox{\plotpoint}}
\put(631.04,1205.65){\usebox{\plotpoint}}
\put(650.83,1211.85){\usebox{\plotpoint}}
\put(670.33,1218.94){\usebox{\plotpoint}}
\put(689.80,1226.12){\usebox{\plotpoint}}
\put(709.22,1233.44){\usebox{\plotpoint}}
\put(728.42,1241.28){\usebox{\plotpoint}}
\put(747.88,1248.50){\usebox{\plotpoint}}
\put(766.98,1256.59){\usebox{\plotpoint}}
\put(786.06,1264.75){\usebox{\plotpoint}}
\put(805.23,1272.65){\usebox{\plotpoint}}
\put(824.47,1280.40){\usebox{\plotpoint}}
\put(843.29,1289.14){\usebox{\plotpoint}}
\put(862.08,1297.95){\usebox{\plotpoint}}
\put(880.98,1306.54){\usebox{\plotpoint}}
\put(899.87,1315.12){\usebox{\plotpoint}}
\put(918.77,1323.71){\usebox{\plotpoint}}
\put(937.43,1332.78){\usebox{\plotpoint}}
\put(955.97,1342.08){\usebox{\plotpoint}}
\put(974.66,1351.09){\usebox{\plotpoint}}
\put(993.27,1360.24){\usebox{\plotpoint}}
\put(1011.67,1369.84){\usebox{\plotpoint}}
\put(1029.90,1379.76){\usebox{\plotpoint}}
\put(1048.51,1388.92){\usebox{\plotpoint}}
\put(1066.73,1398.86){\usebox{\plotpoint}}
\put(1085.18,1408.31){\usebox{\plotpoint}}
\put(1103.33,1418.36){\usebox{\plotpoint}}
\put(1121.55,1428.30){\usebox{\plotpoint}}
\put(1139.77,1438.24){\usebox{\plotpoint}}
\put(1157.90,1448.34){\usebox{\plotpoint}}
\put(1175.98,1458.53){\usebox{\plotpoint}}
\put(1194.20,1468.47){\usebox{\plotpoint}}
\put(1211.97,1479.17){\usebox{\plotpoint}}
\put(1230.19,1489.11){\usebox{\plotpoint}}
\put(1247.86,1500.00){\usebox{\plotpoint}}
\put(1265.80,1510.42){\usebox{\plotpoint}}
\put(1283.73,1520.85){\usebox{\plotpoint}}
\put(1301.68,1531.25){\usebox{\plotpoint}}
\put(1319.35,1542.13){\usebox{\plotpoint}}
\put(1337.26,1552.60){\usebox{\plotpoint}}
\put(1354.80,1563.69){\usebox{\plotpoint}}
\put(1372.74,1574.11){\usebox{\plotpoint}}
\put(1390.01,1585.61){\usebox{\plotpoint}}
\put(1407.89,1596.11){\usebox{\plotpoint}}
\put(1425,1607){\usebox{\plotpoint}}
\end{picture}

\end	{center}
\vskip 0.15in
\caption{The masses of the lightest, $\bullet$, and first
excited, $\circ$, $k=1$ flux loops that wind
around a spatial torus of length $l$ in the SU(4) 
calculation at $\beta=10.9$. The solid line is the
best fit with a bosonic string correction,
as in eqn(\ref{eqn_corr}) with $c=1$. The dotted
lines are the predictions of the Nambu-Goto string action,
as in eqn(\ref{eqn_NGE}). The dynamical lower bound on the 
string length is $l_{min} = 1/aT_c \simeq 6.66$.}
\label{fig_K1n4}
\end 	{figure}
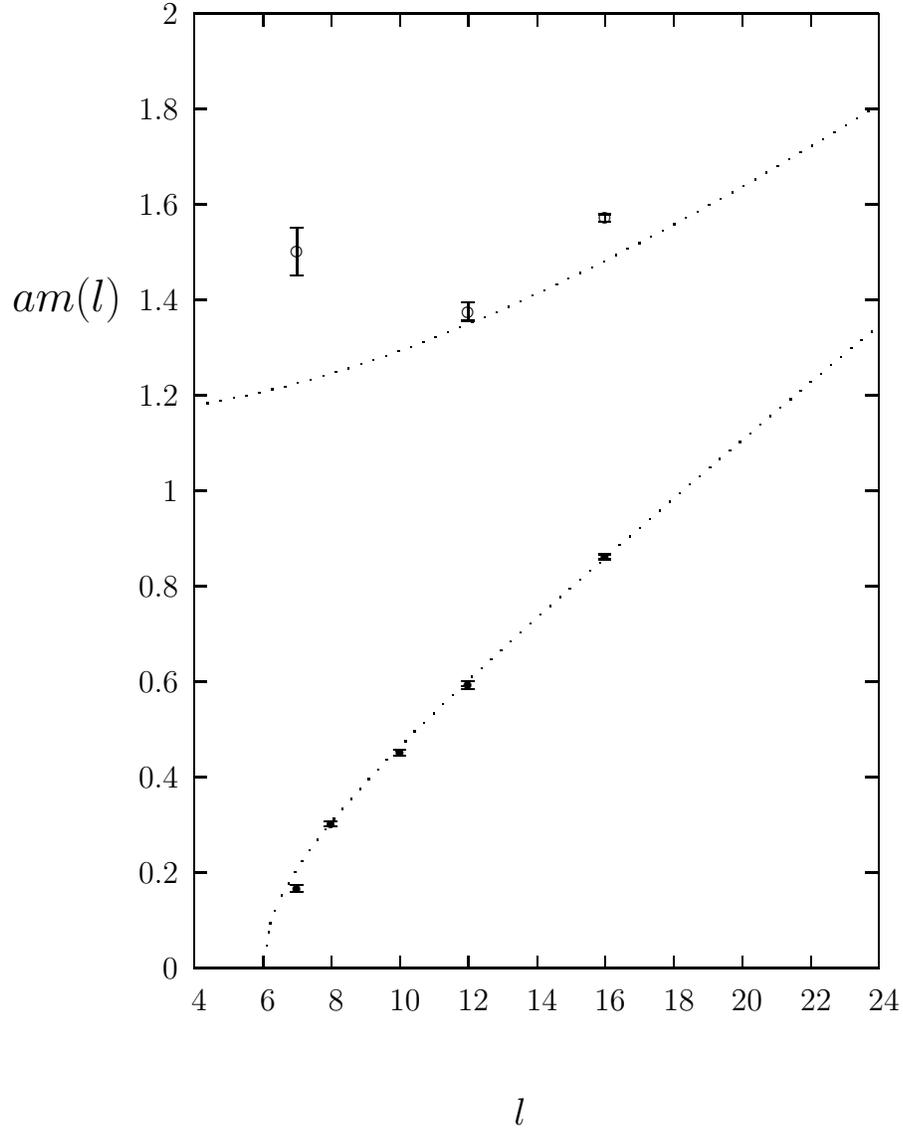

In addition to the above SU(6) calculation we have also 
performed, for comparison, a less extensive and less accurate
SU(4) calculation, as listed in Table~\ref{table_su4}. A fit with
a $O(1/l)$ bosonic string correction plus an extra $O(1/l^3)$ 
correction gives
\begin{equation}
aE_0(l)
= 
0.05798(32) l - \frac{\pi}{3l}-\frac{17.2(4.8)}{l^3}
\quad : \quad l \geq 8 \ \ , \chi^2_{df} = 1.0 
\label{eqn_fitl4k1n4} 
\end{equation}
As with SU(6) the coefficient of the  $O(1/l^3)$ is of the
expected order of magnitude, although it is no longer possible
to include the smallest value of $l$ in the fit. 
In Fig.\ref{fig_K1n4} we plot
the lightest and first excited $k=1$ flux loop masses
and compare to the Nambu-Goto string predictions. As we
remarked earlier, the best Nambu-Goto fit is considerably
worse than for SU(6). It is thus possible that such a fit
will become acceptable in the $N=\infty$ limit.

Fitting the $k= 2$ string masses with an $O(1/l)$ bosonic string 
correction and an additional $O(1/l^3)$ term, we obtain
as acceptable fits,
\begin{equation}
aE^{k=2}_0(l)
= 
\begin{cases} 
0.07963(53) l - {\pi}/{3l}+{1.7(21.2)}/{l^3}
&:\quad l \geq 10 \ \ , \ SU(4) \\
0.10739(75) l - {\pi}/{3l}-{65.7(10.6)}/{l^3}
&:\quad l \geq 8 \ \ \ ,  \ SU(6)
\end{cases}
\label{eqn_fitl4k} 
\end{equation}
The fact that the $O(1/l^3)$ coefficient is much larger 
for $N=6$ than for $N=4$ presumably encodes the fact that
the minimum string length at which the L\"uscher correction 
becomes a good approximation is larger for larger $N$.
(And the fact that it is much larger than for $k=1$ in
eqn(\ref{eqn_fitl4k1n6}) suggests it grows with $k$.)
Together with eqns(\ref{eqn_fitl4k1n6},\ref{eqn_fitl4k1n4}),
these lead to
\begin{equation}
\frac{\sigma_k}{\sigma}
\stackrel{a\simeq 0.15/T_c}{=} 
\begin{cases} 
1.373(12)
&:\quad SU(4) \\
1.689(13)
&:\quad SU(6) 
\end{cases}
\label{eqn_k2k1} 
\end{equation}
These ratios have very small systematic and statistical errors
and show quite clearly that at this fixed but small value of the 
lattice spacing these string tensions lie in
between the Casimir scaling 
\cite{CS}
and `MQCD' conjectures 
\cite{MQCD}
with which they are usually compared
\cite{blmt-string01,pisa-kstring,blmtuw-glue04}.

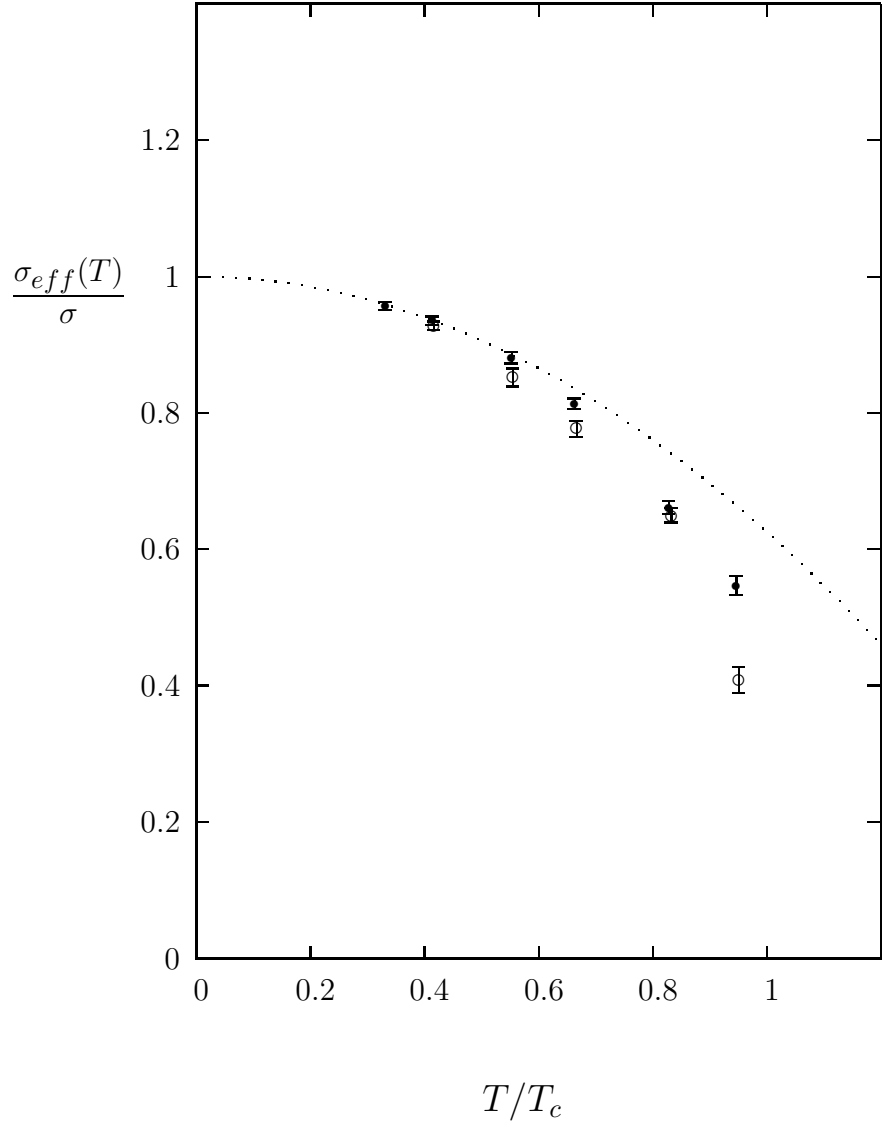
\begin	{figure}[p]
\begin	{center}
\leavevmode
\setlength{\unitlength}{0.240900pt}
\ifx\plotpoint\undefined\newsavebox{\plotpoint}\fi
\sbox{\plotpoint}{\rule[-0.200pt]{0.400pt}{0.400pt}}%
\begin{picture}(1500,1800)(0,0)
\font\gnuplot=cmr10 at 12pt
\gnuplot
\sbox{\plotpoint}{\rule[-0.200pt]{0.400pt}{0.400pt}}%
\put(350.0,250.0){\rule[-0.200pt]{4.818pt}{0.400pt}}
\put(325,250){\makebox(0,0)[r]{\ \ {$0$}}}
\put(1405.0,250.0){\rule[-0.200pt]{4.818pt}{0.400pt}}
\put(350.0,464.0){\rule[-0.200pt]{4.818pt}{0.400pt}}
\put(325,464){\makebox(0,0)[r]{\ \ {$0.2$}}}
\put(1405.0,464.0){\rule[-0.200pt]{4.818pt}{0.400pt}}
\put(350.0,679.0){\rule[-0.200pt]{4.818pt}{0.400pt}}
\put(325,679){\makebox(0,0)[r]{\ \ {$0.4$}}}
\put(1405.0,679.0){\rule[-0.200pt]{4.818pt}{0.400pt}}
\put(350.0,893.0){\rule[-0.200pt]{4.818pt}{0.400pt}}
\put(325,893){\makebox(0,0)[r]{\ \ {$0.6$}}}
\put(1405.0,893.0){\rule[-0.200pt]{4.818pt}{0.400pt}}
\put(350.0,1107.0){\rule[-0.200pt]{4.818pt}{0.400pt}}
\put(325,1107){\makebox(0,0)[r]{\ \ {$0.8$}}}
\put(1405.0,1107.0){\rule[-0.200pt]{4.818pt}{0.400pt}}
\put(350.0,1321.0){\rule[-0.200pt]{4.818pt}{0.400pt}}
\put(325,1321){\makebox(0,0)[r]{\ \ {$1$}}}
\put(1405.0,1321.0){\rule[-0.200pt]{4.818pt}{0.400pt}}
\put(350.0,1536.0){\rule[-0.200pt]{4.818pt}{0.400pt}}
\put(325,1536){\makebox(0,0)[r]{\ \ {$1.2$}}}
\put(1405.0,1536.0){\rule[-0.200pt]{4.818pt}{0.400pt}}
\put(350.0,250.0){\rule[-0.200pt]{0.400pt}{4.818pt}}
\put(350,200){\makebox(0,0){\ {$0$}}}
\put(350.0,1730.0){\rule[-0.200pt]{0.400pt}{4.818pt}}
\put(529.0,250.0){\rule[-0.200pt]{0.400pt}{4.818pt}}
\put(529,200){\makebox(0,0){\ {$0.2$}}}
\put(529.0,1730.0){\rule[-0.200pt]{0.400pt}{4.818pt}}
\put(708.0,250.0){\rule[-0.200pt]{0.400pt}{4.818pt}}
\put(708,200){\makebox(0,0){\ {$0.4$}}}
\put(708.0,1730.0){\rule[-0.200pt]{0.400pt}{4.818pt}}
\put(888.0,250.0){\rule[-0.200pt]{0.400pt}{4.818pt}}
\put(888,200){\makebox(0,0){\ {$0.6$}}}
\put(888.0,1730.0){\rule[-0.200pt]{0.400pt}{4.818pt}}
\put(1067.0,250.0){\rule[-0.200pt]{0.400pt}{4.818pt}}
\put(1067,200){\makebox(0,0){\ {$0.8$}}}
\put(1067.0,1730.0){\rule[-0.200pt]{0.400pt}{4.818pt}}
\put(1246.0,250.0){\rule[-0.200pt]{0.400pt}{4.818pt}}
\put(1246,200){\makebox(0,0){\ {$1$}}}
\put(1246.0,1730.0){\rule[-0.200pt]{0.400pt}{4.818pt}}
\put(350.0,250.0){\rule[-0.200pt]{258.967pt}{0.400pt}}
\put(1425.0,250.0){\rule[-0.200pt]{0.400pt}{361.350pt}}
\put(350.0,1750.0){\rule[-0.200pt]{258.967pt}{0.400pt}}
\put(150,1300){\makebox(0,0){\Large{$\frac{\sigma_{eff}(T)}{\sigma}$}}}
\put(862,25){\makebox(0,0){\large{$T/T_c$}}}
\put(350.0,250.0){\rule[-0.200pt]{0.400pt}{361.350pt}}
\put(1198.0,821.0){\rule[-0.200pt]{0.400pt}{7.227pt}}
\put(1188.0,821.0){\rule[-0.200pt]{4.818pt}{0.400pt}}
\put(1188.0,851.0){\rule[-0.200pt]{4.818pt}{0.400pt}}
\put(1092.0,948.0){\rule[-0.200pt]{0.400pt}{5.059pt}}
\put(1082.0,948.0){\rule[-0.200pt]{4.818pt}{0.400pt}}
\put(1082.0,969.0){\rule[-0.200pt]{4.818pt}{0.400pt}}
\put(944.0,1113.0){\rule[-0.200pt]{0.400pt}{4.095pt}}
\put(934.0,1113.0){\rule[-0.200pt]{4.818pt}{0.400pt}}
\put(934.0,1130.0){\rule[-0.200pt]{4.818pt}{0.400pt}}
\put(845.0,1185.0){\rule[-0.200pt]{0.400pt}{4.336pt}}
\put(835.0,1185.0){\rule[-0.200pt]{4.818pt}{0.400pt}}
\put(835.0,1203.0){\rule[-0.200pt]{4.818pt}{0.400pt}}
\put(721.0,1246.0){\rule[-0.200pt]{0.400pt}{3.132pt}}
\put(711.0,1246.0){\rule[-0.200pt]{4.818pt}{0.400pt}}
\put(711.0,1259.0){\rule[-0.200pt]{4.818pt}{0.400pt}}
\put(647.0,1269.0){\rule[-0.200pt]{0.400pt}{3.132pt}}
\put(637.0,1269.0){\rule[-0.200pt]{4.818pt}{0.400pt}}
\put(1198,836){\circle*{12}}
\put(1092,958){\circle*{12}}
\put(944,1121){\circle*{12}}
\put(845,1194){\circle*{12}}
\put(721,1253){\circle*{12}}
\put(647,1275){\circle*{12}}
\put(637.0,1282.0){\rule[-0.200pt]{4.818pt}{0.400pt}}
\put(1202.0,667.0){\rule[-0.200pt]{0.400pt}{9.877pt}}
\put(1192.0,667.0){\rule[-0.200pt]{4.818pt}{0.400pt}}
\put(1192.0,708.0){\rule[-0.200pt]{4.818pt}{0.400pt}}
\put(1096.0,935.0){\rule[-0.200pt]{0.400pt}{5.541pt}}
\put(1086.0,935.0){\rule[-0.200pt]{4.818pt}{0.400pt}}
\put(1086.0,958.0){\rule[-0.200pt]{4.818pt}{0.400pt}}
\put(947.0,1070.0){\rule[-0.200pt]{0.400pt}{6.022pt}}
\put(937.0,1070.0){\rule[-0.200pt]{4.818pt}{0.400pt}}
\put(937.0,1095.0){\rule[-0.200pt]{4.818pt}{0.400pt}}
\put(847.0,1149.0){\rule[-0.200pt]{0.400pt}{6.745pt}}
\put(837.0,1149.0){\rule[-0.200pt]{4.818pt}{0.400pt}}
\put(837.0,1177.0){\rule[-0.200pt]{4.818pt}{0.400pt}}
\put(723.0,1238.0){\rule[-0.200pt]{0.400pt}{3.132pt}}
\put(713.0,1238.0){\rule[-0.200pt]{4.818pt}{0.400pt}}
\put(1202,687){\circle{18}}
\put(1096,946){\circle{18}}
\put(947,1083){\circle{18}}
\put(847,1163){\circle{18}}
\put(723,1244){\circle{18}}
\put(713.0,1251.0){\rule[-0.200pt]{4.818pt}{0.400pt}}
\put(350,1321){\usebox{\plotpoint}}
\put(350.00,1321.00){\usebox{\plotpoint}}
\put(370.76,1321.00){\usebox{\plotpoint}}
\put(391.47,1320.15){\usebox{\plotpoint}}
\put(412.18,1319.26){\usebox{\plotpoint}}
\put(432.90,1318.37){\usebox{\plotpoint}}
\put(453.57,1316.49){\usebox{\plotpoint}}
\put(474.08,1313.54){\usebox{\plotpoint}}
\put(494.62,1310.67){\usebox{\plotpoint}}
\put(515.13,1307.61){\usebox{\plotpoint}}
\put(535.55,1303.89){\usebox{\plotpoint}}
\put(555.94,1300.01){\usebox{\plotpoint}}
\put(576.18,1295.50){\usebox{\plotpoint}}
\put(596.41,1290.98){\usebox{\plotpoint}}
\put(616.53,1285.89){\usebox{\plotpoint}}
\put(636.50,1280.36){\usebox{\plotpoint}}
\put(656.35,1274.36){\usebox{\plotpoint}}
\put(676.09,1267.98){\usebox{\plotpoint}}
\put(695.77,1261.49){\usebox{\plotpoint}}
\put(715.26,1254.36){\usebox{\plotpoint}}
\put(734.76,1247.27){\usebox{\plotpoint}}
\put(754.20,1240.00){\usebox{\plotpoint}}
\put(773.29,1231.87){\usebox{\plotpoint}}
\put(792.18,1223.28){\usebox{\plotpoint}}
\put(811.08,1214.69){\usebox{\plotpoint}}
\put(829.97,1206.10){\usebox{\plotpoint}}
\put(848.30,1196.42){\usebox{\plotpoint}}
\put(866.89,1187.24){\usebox{\plotpoint}}
\put(885.51,1178.09){\usebox{\plotpoint}}
\put(903.73,1168.15){\usebox{\plotpoint}}
\put(921.95,1158.21){\usebox{\plotpoint}}
\put(939.45,1147.12){\usebox{\plotpoint}}
\put(957.26,1136.47){\usebox{\plotpoint}}
\put(975.20,1126.06){\usebox{\plotpoint}}
\put(992.71,1114.91){\usebox{\plotpoint}}
\put(1009.98,1103.41){\usebox{\plotpoint}}
\put(1026.96,1091.48){\usebox{\plotpoint}}
\put(1044.47,1080.34){\usebox{\plotpoint}}
\put(1061.28,1068.16){\usebox{\plotpoint}}
\put(1078.06,1055.95){\usebox{\plotpoint}}
\put(1094.49,1043.28){\usebox{\plotpoint}}
\put(1111.28,1031.07){\usebox{\plotpoint}}
\put(1127.76,1018.47){\usebox{\plotpoint}}
\put(1144.30,1005.94){\usebox{\plotpoint}}
\put(1160.11,992.50){\usebox{\plotpoint}}
\put(1176.01,979.17){\usebox{\plotpoint}}
\put(1192.07,966.03){\usebox{\plotpoint}}
\put(1208.13,952.88){\usebox{\plotpoint}}
\put(1223.70,939.16){\usebox{\plotpoint}}
\put(1238.92,925.08){\usebox{\plotpoint}}
\put(1254.71,911.63){\usebox{\plotpoint}}
\put(1270.07,897.67){\usebox{\plotpoint}}
\put(1285.36,883.64){\usebox{\plotpoint}}
\put(1300.27,869.21){\usebox{\plotpoint}}
\put(1315.20,854.80){\usebox{\plotpoint}}
\put(1329.88,840.12){\usebox{\plotpoint}}
\put(1344.86,825.77){\usebox{\plotpoint}}
\put(1359.72,811.28){\usebox{\plotpoint}}
\put(1374.24,796.46){\usebox{\plotpoint}}
\put(1388.24,781.14){\usebox{\plotpoint}}
\put(1402.72,766.28){\usebox{\plotpoint}}
\put(1416.89,751.11){\usebox{\plotpoint}}
\put(1425,743){\usebox{\plotpoint}}
\end{picture}

\end	{center}
\vskip 0.15in
\caption{The finite temperature effective string tension 
$\sigma_{eff}(T)$ plotted as a function of $T/T_c$ for 
SU(6), $\bullet$, and SU(4), $\circ$, at a fixed
lattice spacing $a \simeq 0.15/T_c$.}
\label{fig_KeffT}
\end 	{figure}

As an aside, we note that
the above calculations also tell us about how the effective
string tension, $\sigma_{eff}$, varies with temperature $T$. 
Let us relabel 
our Euclidean axes so that our $x$ direction is the $t$
direction. Then our $k=1$ Wilson line operator is just the 
world line of a static fundamental source and if we define
$\sigma_{eff}$ to be the coefficient of the linear piece
of the free energy of two such static sources then it
is clear that
\cite{blmtuw-T03,blmtuw-T04}
\begin{equation}
\sigma_{eff}(T)
\stackrel{l=1/T}{=} 
\frac{E_0(l)}{l}
\stackrel{l=1/T}{=} 
\sigma - \frac{\pi}{3}T^2 + O(T^4)
\label{eqn_Keff} 
\end{equation}
if the string is in the simplest bosonic universality class.
In Fig.\ref{fig_KeffT} we plot the values of $\sigma_{eff}(T)$
against $T/T_c$ for both SU(4) and SU(6). There is a hint 
that as $N$ increases the effective string tension has
smaller corrections at all $T$.

\section{Conclusions}
\label{section_conclusions}

In our calculation in the SU(6) gauge theory we have obtained 
good evidence for linear confinement of flux loops that
are up to $l \simeq 5/\surd\sigma$ long. This is
long enough in physical units (it corresponds to
$\sim 2.5{\mathrm fm}$ in QCD) for us to see it as good
evidence for exact linear confinement. The calculation is
at a fixed lattice spacing but this is small enough,
$a \sim 1/4\surd\sigma$, that we are confident that
we are seeing continuum physics. Finally $N=6$ is large
enough that we see this is as good evidence for linear
confinement in the $N\to\infty$ limit.

For strings in the fundamental representation our SU(6)
calculation shows that the leading $O(1/l)$ correction
at large $l$ has the coefficient one expects if the
effective string theory belongs to the simplest bosonic
string universality class. This coefficient is determined
accurately enough for this result to be significant.

These calculations were performed with care to control and
quantify the main systematic errors. We found that string
tensions could be reliably calculated on $L^4$ lattices
once $L \geq 3/a\surd\sigma$. This fortunately coincides
with the conventional choice of lattice size for such
calculations
\cite{blmt-string01}. 
For flux loops in higher representations our results do not
pin down the string corrections very accurately, but
they are consistent with a bosonic string correction
dominating at large $l$. There is, however, evidence
that this occurs at larger $l$ than for the $k=1$ string, the
more so the larger $N$, and that the effective string coefficient
at smaller $l$ is larger for larger $k$. This fits in with the 
picture of a $k$ string as being $k$ loosely bound fundamental
strings with the L\"uscher correction inducing an 
effective repulsive interaction. It also implies that
past calculations of $k$-string tensions may have suffered from
a small but systematic downward bias. The ratio of the $k=2$ and
$k=1$ string tensions that we obtain in this paper lies 
more-or-less mid-way between the Casimir Scaling and
the `MQCD' motivated trigonometric formula.

Remarkably, we found that using only the $O(1/l)$ bosonic string 
correction with an additional  $O(1/l^3)$ term, we could describe 
the (fundamental) string energy all the way down to the minimum string 
length, $l\simeq 1/T_c$ where $T_c$ is the deconfining temperature.
The fitted coefficient of the  $O(1/l^3)$ term turns out to 
be $O(1/a^2\sigma)$ just as one would expect on general grounds. 

The string energy is also found to be close to the prediction of the
simplest string theory, that governed by the Nambu-Goto action,
and the comparison between SU(4) and SU(6) leaves open
the possibility that for SU($\infty$) the effective string
theory is indeed governed by that action -- although this would
be hard to understand on theoretical grounds
\cite{largeN-string,largeN-ads}. 
Our calculation of the energy of the first excited string
and our discussion of string deconfinement, reinforces the
conclusion that whatever effective string action governs the
$N=\infty$ theory, it must be `quite close' to Nambu-Goto.
  
It would be useful to have, say, SU(4) and SU(8) calculations
that are as precise as our SU(6) ones, so that we could 
directly extrapolate the string energy to $N=\infty$ as a function
of $l$, all the way down down to the minimum length,
$l\simeq 1/T_c$, dictated by `spatial deconfinement'. In fact, as we 
remarked earlier, one can use the hysteresis of this first order phase 
transition to go to lower $l$. The aim would be to go all the way 
down to the second order string condensation transition where the
string mass vanishes. Although it appears that the hysteresis
is not wide enough for this to be possible in SU(8), there is
good reason to believe it will be possible for not very much larger 
values of $N$. The low-$l$ behaviour of the string energy
together with the excited string spectrum, particularly for
states whose energies do not approach the ground state energy
at large $l$, should provide useful insights into the
detailed nature of the effective string theory for, and the
dual string theory to, the SU($\infty$) gauge theory.

\section*{Acknowledgements}

Our lattice calculations were carried out on PPARC 
and EPSRC funded computers in Oxford Theoretical 
Physics, and on a desktop funded by All Souls College.
During part of this work,
MT participated in the `QCD and String Theory' Workshop 
at the KITP UCSB and acknowledges very useful discussions
with many of the participants.

\vfill\eject

\begin{table}
\begin{center}
\begin{tabular}{|c|c|c|c|c|c|c|}\hline
\multicolumn{7}{|c|}{ SU(6) } \\ \hline
 $l_x$ & $l_{y,z}$ & $l_t$ & $am(k=1)$ &  $am^\star(k=1)$ & $am(k=2)$ & $am(k=3)$ \\ \hline
7  & 12 & 36 & 0.19(3)    &           & 0.24(11)  & 0.22(16) \\
7  & 16 & 40 & 0.242(6)   & 1.635(66) & 0.488(16) & 0.624(23) \\ \hline
8  & 8  & 14 & 0.295(6)   &           & 0.414(17) & 0.455(24) \\
8  & 8  & 30 & 0.302(8)   &           & 0.480(19) & 0.534(28) \\
8  & 10 & 30 & 0.343(8)   & 1.244(31) & 0.582(19) & 0.703(41) \\
8  & 12 & 30 & 0.3343(50) & 1.234(17) & 0.596(13) & 0.710(18) \\ \hline
10 & 10 & 10 & 0.494(6)   &           & 0.800(17) & 0.973(31) \\
10 & 10 & 16 & 0.505(9)   & 1.392(21) & 0.863(19) & 1.039(27) \\
10 & 12 & 16 & 0.5141(50) & 1.366(19) & 0.907(10) & 1.052(18) \\ \hline
12 & 12 & 12 & 0.669(6)   & 1.497(25) & 1.148(16) & 1.354(29) \\
12 & 14 & 14 & 0.6683(55) & 1.505(24) & 1.180(13) & 1.422(23) \\ \hline
16 & 16 & 16 & 0.9465(59) & 1.638(22) & 1.625(20) & 1.878(45) \\
16$^{\star}$ & 16 & 24 & 0.930(10) & 1.627(17) & 1.634(19) & 1.844(60) \\ \hline
20 & 20 & 24 & 1.210(7)   &  &   & 2.24(8) \\ 
20$^{\star}$ & 20 & 24 & 1.242(15)   & 1.800(51) & 2.077(67) &   \\ \hline
\end{tabular}
\caption{\label{table_su6} }
\end{center}
\end{table}

\begin{table}
\begin{center}
\begin{tabular}{|c||c|c||c|c||c|c|}\hline
\multicolumn{7}{|c|}{ SU(6) : $\sigma l - c\pi/3l$ fits} \\ \hline
 $l_x\geq$ & $c(k=1)$ & $\chi^2/n_{df}$ & $c(k=2)$ & $\chi^2/n_{df}$ & $c(k=3)$ & $\chi^2/n_{df}$  \\ \hline
7  & 1.36(4)  & 3.8 & 2.02(12) &  2.1 & 1.96(19) & 5.4 \\
8  & 1.31(5)  & 4.0 & 2.11(15) &  2.3 & 2.26(25) & 5.7 \\
10 & 1.16(8)  & 0.9 & 1.73(22) &  0.4 & 1.48(42) & 5.7 \\
12 & 1.09(10) & 0.7 & 1.45(40) &  0.0 & -0.38(73) & 0.8 \\ \hline
\end{tabular}
\caption{\label{table_su6fit} }
\end{center}
\end{table}

\begin{table}
\begin{center}
\begin{tabular}{|c|c|c|c|c|c|}\hline
\multicolumn{6}{|c|}{ SU(4) } \\ \hline
 $l_x$ & $l_{y,z}$ & $l_t$ & $am(k=1)$ & $am^\star(k=1)$ & $am(k=2)$  \\ \hline
7  & 20 & 60 & 0.1655(75) & 1.50(5)   & 0.261(16) \\
8  & 16 & 24 & 0.3015(50) &           & 0.470(15) \\
10 & 16 & 24 & 0.4506(69) &           & 0.698(13) \\
12 & 16 & 24 & 0.5928(88) & 1.375(19) & 0.856(15) \\
16 & 16 & 24 & 0.8610(53) & 1.571(8)  & 1.2102(63) \\ \hline
\end{tabular}
\caption{\label{table_su4} }
\end{center}
\end{table}

\end{document}